\documentclass[acmtog, screen, nonacm]{acmart}
\AtBeginDocument{%
  \providecommand\BibTeX{{%
    \normalfont B\kern-0.5em{\scshape i\kern-0.25em b}\kern-0.8em\TeX}}}

\AtBeginDocument{%
  \providecommand\BibTeX{{%
    \normalfont B\kern-0.5em{\scshape i\kern-0.25em b}\kern-0.8em\TeX}}}

\acmSubmissionID{379}
\copyrightyear{2024}

\usepackage[skip=0.5em,textfont=it,labelfont=bf]{caption}
\usepackage{standalone}
\usepackage{wrapfig}
\usepackage{tikz}
\usepackage{graphicx}
\usepackage{subcaption}
\usepackage{amsmath,bm}
\usepackage{booktabs}
\usepackage{algorithm}
\usepackage{algpseudocode}
\usepackage{color}
\usepackage{soul}
\usepackage{cancel}
\usepackage{multicol}

\usepackage[normalem]{ulem}
\usepackage{xcolor}

\newcommand{\highlight}[1]{\textcolor{black}{#1}}

\let\oldnl\nl
\newcommand{\nonl}{\renewcommand{\nl}{\let\nl\oldnl}}
\newlength\savedwidth
\newcommand\whline[1]{\noalign{\global\savedwidth\arrayrulewidth
                               \global\arrayrulewidth #1} 
                      \hline
                      \noalign{\global\arrayrulewidth\savedwidth}}

\begin{document}

\title{\highlight{Volumetric Homogenization} for Knitwear Simulation}

\author{Chun Yuan}{
\authornote{joint first authors}
\email{yuanchunisme@gmail.com}
\affiliation{%
  \institution{University of Utah}
  \country{USA}
}

\author{Haoyang Shi}
\authornotemark[1]
\email{u1431587@utah.edu}
\affiliation{%
  \institution{University of Utah}
  \country{USA}}

\author{Lei Lan}
\email{lanlei.virhum@gmail.com}
\affiliation{%
  \institution{University of Utah}
  \country{USA}
}

\author{Yuxing Qiu}
\email{yuxqiu@gmail.com}
\affiliation{%
 \institution{LightSpeed Studios}
 \city{Palo Alto}
 \country{USA}}

\author{Cem Yuksel}
\email{cem@cemyuksel.com}
\affiliation{%
  \institution{University of Utah}
  \city{Salt Lake City}
  \country{USA}}

\author{Huamin Wang}
\email{wanghmin@gmail.com}
\affiliation{%
  \institution{Style3D Research}
  \city{Galena}
  \country{USA}}

\author{Chenfanfu Jiang}
\email{chenfanfu.jiang@gmail.com}
\affiliation{%
  \institution{UCLA}
  \city{Los Angeles}
  \country{USA}}

\author{Kui Wu}
\email{walker.kui.wu@gmail.com}
\affiliation{%
  \institution{LightSpeed Studios}
  \city{Los Angeles}
  \country{USA}}

\author{Yin Yang}
\email{yangzzzy@gmail.com}
\affiliation{%
  \institution{University of Utah}
  \city{Salt Lake City}
  \country{USA}}


\begin{abstract}
This paper presents \highlight{volumetric homogenization}, a spatially varying homogenization scheme for knitwear simulation. We are motivated by the observation that macro-scale fabric dynamics is strongly correlated with its underlying knitting patterns. Therefore, homogenization towards a single material is less effective when the knitting is complex and non-repetitive. Our method tackles this challenge by homogenizing the yarn-level material locally at volumetric elements. Assigning a virtual volume of a knitting structure enables us to model bending and twisting effects via a simple volume-preserving penalty and thus effectively alleviates the material nonlinearity. We employ an adjoint Gauss-Newton formulation\highlight{\cite{Zehnder2021}} to battle the dimensionality challenge of such per-element material optimization. This intuitive material model makes the forward simulation GPU-friendly. To this end, our pipeline also equips a novel domain-decomposed subspace solver crafted for GPU projective dynamics, which makes our simulator hundreds of times faster than the yarn-level simulator. Experiments validate the capability and effectiveness of \highlight{volumetric homogenization}. Our method produces realistic animations of knitwear matching the quality of full-scale yarn-level simulations. It is also orders of magnitude faster than existing homogenization techniques in both the training and simulation stages.
\end{abstract}




\begin{teaserfigure}
  \includegraphics[width=\textwidth]{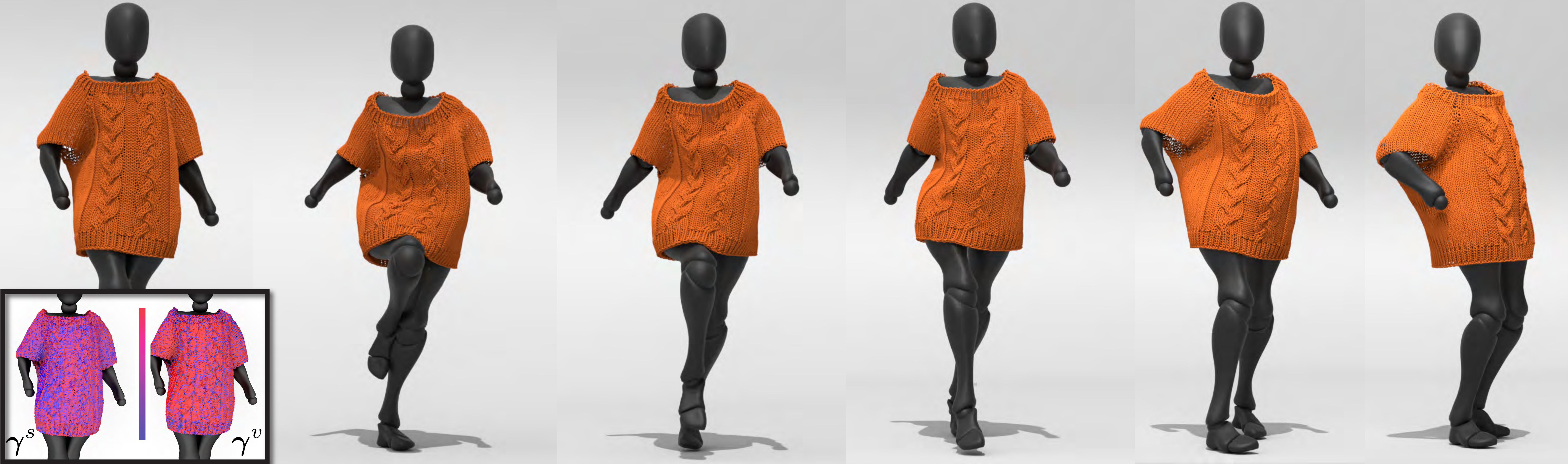}
  \caption{\textbf{Yangge dance.}~~We propose a \highlight{volumetric homogenization} algorithm for knitwear simulation. Our pipeline takes full-scale yarn-level simulation observations as input and learns a homogenization out of the input poses. We name this procedure \highlight{volumetric-homogenization} because we do not homogenize yarn-level dynamics to a codimensional sheet/shell domain but to a volumetric enclosure. Doing so allows us to use an intuitive hyperelastic material model with a volume-preserving constraint to capture nonlinear behaviors like bending and twisting. \highlight{volumetric homogenization} varies spatially. Each volumetric element has its own material parameter, which substantially enhances the expressivity of our model. \highlight{This is a challenging problem, we utilize the adjoint Gauss-Newton method\cite{Zehnder2021} to enhance the convergence of the adjoint method. Additionally, we propose harmonic initialization, sample-based optimization, and a novel domain-decomposed projective dynamics solver, which significantly accelerates the entire optimization process.} In the teaser figure, we show snapshots of an animated virtual character performing a Yangge dance, where we simulate a homogenized tetrahedron mesh with about $390$K elements that encapsulate a sweater. Cable patterns on the garment contribute to an irregular and sophisticated material distribution (as visualized in the sub-figure). In this example, the mesh-level simulation runs at $604$ ms per frame on average under $\Delta t = 1/150$ sec. The yarn-level sweater model, on the other hand, consists of over $3$M DOFs. Our \highlight{volumetric homogenization} makes the simulation $\sim 100\times$ faster than simulating the sweater at the yarn level.}
  \label{fig:teaser}
  \Description{}
\end{teaserfigure}

\maketitle

\section{Introduction}
Creating realistic garment animation is a core problem of computer graphics. A common practice is to leverage a mass-spring system~\cite{provot1995deformation,liu2013fast} or a triangle mesh~\cite{baraff1998large} to simulate the cloth motion as a thin membrane. The resultant cloth dynamics are controlled by a few macroscopic (triangle-level) material parameters like bending and stretching stiffness. On the other hand, knitwear like sweaters, scarves, and cardigans that are fabricated by interlocking thick yarn threads exhibit ``non-rubbery'' behaviors due to intricate yarn-level interactions. The intriguing mechanical responses of knitted fabrics are beyond the expressivity of an elastic sheet and require dedicated algorithms to simulate microscopic dynamics, such as yarn-level simulation (YLS)~\cite{kaldor2008simulating}. While YLS is able to produce high-fidelity dynamics, it is also known to be computationally expensive as one needs a large number of degrees of freedom (DOFs) to capture the motions and interactions of individual yarn threads. 

Homogenized yarn-level cloth (HYLC)~\cite{sperl2020homogenized} aims to improve the YLS efficiency by regressing a nonlinear hyperelastic thin shell model based on quasi-static YLS responses so that the triangle-level simulation reflects the dominant effects of the knits. It is based on the classic theory of computational homogenization~\cite{geers2010multi} that uniform macro-scale material properties can be extracted out of micro-scale structural variations. This applies to the knit structure with relatively simple and periodic patterns, as the inconsistency between macroscopic and microscopic motion is believed to be averaged out. However, knit sweaters often feature intricate patterns and complex yarn structures across their front panels, and the spatial disparity makes the traditional homogenization technique cumbersome. Moreover, determining the appropriate size for the representative volume element (RVE)~\cite{de2015rve,liu2016discrete} is not straightforward in such cases. Another limitation (maybe more as a natural consequence) of HYLC is its reliance on a highly nonlinear material model to capture the non-smooth yarn movements at micro scales, leading to potential numerical instability, such as energy singularity and Hessians that are not positive definite. Therefore, HYLC simulation uses a (very) conservative time step size, resulting in more steps to compute, which undermines 
its original purpose of computational efficiency.

Simply increasing the material complexity in the homogenization, e.g., using spline-based strain-stress model~\cite{xu2015nonlinear,sperl2020homogenized} or even neural-network-based materials~\cite{fengneural2024}, is less helpful to capture the dynamics of complicated interlocked yarn structure with large variations and often induce numerical instability (e.g. see~\citet{wang2023stable}).

We propose an entirely
different perspective for efficient simulation of knits that can even handle non-repeating knitting patterns. In contrast to traditional material homogenization methods or HYLC, we introduce a novel approach: \highlight{volumemetric homogenization}. We focus on studying the \emph{heterogeneity}, i.e., the spatial variation of material parameters while keeping the constitutional relation simple. In particular, we incorporate a virtual volume representing the yarn material and exploit spatially varying volume-preserving penalties to capture desired knit characteristics, such as stretching, shearing, and bending, without over-complicating the material model.



We discretize the fabric at the yarn level and embed it within a volumetric mesh. This mesh is of high resolution to capture the local structural characteristics of yarn loops. 
We learn a macro-scale material model over this volumetric space, aligning it with dynamic YLS results.
In that respect, our method can also be understood as applying local homogenization at each mesh element. 
Our rationale resembles curve fitting, where using multiple low-degree polynomials (e.g., splines) is often preferred over employing a single high-degree polynomial. Likewise, a composite material model, despite its simplicity, offers a more versatile design space compared to a homogenized material, since each volumetric element possesses independent material freedoms. We \highlight{employ the} adjoint Gauss-Newton method\cite{Zehnder2021} to regress the yarn material via a high-dimension space-time optimization problem. The simplicity of the material model
makes the \highlight{volumetric} homogenization well-suited for efficient forward GPU solvers such as projective dynamics (PD)~\cite{bouaziz2023projective}. To this end, we devise a domain-decomposed multi-level solver for the global stage of PD specially crafted for our runtime pipeline. 

Although simulating over a volume mesh may appear expensive initially, the aforementioned benefits outweigh the overhead of mesh DOFs. As a result, our method is more stable, runs orders of magnitude faster than existing homogenized YLS, and produces realistic animations of knitwear with complex knit patterns.

\section{Related work}\label{sec:related}
There exists a large volume of excellent work on relevant topics of cloth and yarn simulation. Due to the page count limits, this section only briefly discusses some closely relevant prior work.  

\paragraph{Sheet-level cloth model}
Modern fabrics and garments are fabricated through sophisticated workmanship and exhibit intricate material response. Nevertheless, modeling cloth with a sheet-like shell remains prevalent~\cite{choi2005stable,grinspun2003discrete} due to its simplicity and intuitiveness. Mass-spring and/or triangle meshes offer a natural discretization of sheet-based cloth, and its equation of motion can be derived via elastic energies that measure the cloth deformation~\cite{terzopoulos1987elastically}. Implicit integration has become a standard component since the seminal work of \citet{baraff1998large}. Unlike a rubber membrane, cloth fabrics are less stretchable but tend to be bent more easily. This feature inspires various techniques to enforce the inextensibility~\cite{provot1995deformation,goldenthal2007efficient,english2008animating,wang2010multi} with numerical stability. Cloth bending, on the other hand, can be parameterized by the dihedral angle or bending modes~\cite{bridson2005simulation}. It is also possible to exploit
isometric mesh deformations to use the so-called quadratic bending~\cite{Bergou2006,garg2007cubic}, which possesses a constant Hessian. \citet{kim2020finite} reveals the connection between model of \citet{baraff1998large} and the finite element method (FEM)~\cite{bathe2006finite}. To capture detailed wrinkles and folds on the cloth, adaptive remeshing~\cite{narain2013folding,narain2012adaptive} or mixed discretization~\cite{guo2018material,weidner2018eulerian} are proven to be effective. Another important aspect of sheet-level cloth animation is to handle collisions, especially self-collisions~\cite{volino2000implementing,bridson2002robust,baraff2003untangling}. Penalty-based methods~\cite{bouaziz2023projective,wu2020repulsioncloth} are straightforward to implement but often with stability issues. \citet{Provot1997ImpactZone,Harmon2008ImpactZone} introduce approaches that group multiple collisions into ``impact zones'' and treat them as rigid bodies, allowing for some sliding motion. Constraint-based collision handling methods represent contact as constraints based on exact Coulomb friction~\cite{Otaduy2009ImplicitContact, li2018ImplicitFrictional}.

\paragraph{Yarn-level cloth model}
The rapid development of computing hardware makes animated yarn-level fabric feasible. YLS models each yarn thread as an elastic rod~\cite{bergou2010discrete,bergou2008discrete,pai2002strands,spillmann2007corde} and exploits yarn-yarn contact to trigger fabric deformation. This effort is pioneered by \citet{kaldor2008simulating}, where yarn threads are modeled as cubic B-splines. With YLS, different knit structures like garter, rib, and stockinette showcase unique stretching behaviors, which are beyond the expressivity of average sheet-level models. This approach is subsequently accelerated by approximating yarn-yarn collisions using a co-rotated force model~\cite{Kaldor2010adaptivecontact}. Persistent contact~\cite{Cirio2014woven,cirio2015efficient} is an efficient method to handle the interaction among heavily interlaced yarns. This method is further combined with a triangle-based model to create a hybrid system that enhances yarn-level details only in areas of interest~\cite{Casafranca2020mix}. Additionally, \citet{Banderas2020eol} extend the concept of persistent contact to handle stacked fabrics, addressing both intra-fabric and inter-fabric contacts implicitly. 
Computational design and machine fabrication of yarn patterns are also of great interest to various communities. \citet{leaf2018interactive} propose an efficient GPU yarn-level simulator to enable the interactive designing of periodic yarn patches. \citet{yuksel2012stitch} introduce an efficient 3D design and modeling interface at the stitch level, which is later extended for hand knitting~\cite{wu2019knittable}, machine-knitting~\cite{narayanan2019visual}, and enforcing wearability~\cite{wu2021} via cloth simulation.

\paragraph{Data- and learning-based cloth model}
Data-driven or learning-based cloth simulation is considered an effective approach to enhance the realism of continuum shell models. \citet{Wang2011DataDriven} estimate planar and bending stiffness through separate tests, while automatic devices have been developed for acquiring fabric stiffness~\cite{Miguel2012DataDriven} and friction parameters~\cite{Miguel2013DataDriven}. Similar tensile tests have also been utilized by~\citet{Clyde2017DataDriven} to estimate the planar stiffness of woven fabrics. \citet{feng2022learning} use a simulation-in-the-loop framework to estimate the fabric bending stiffness. ClothCap aims to reconstruct multiple garments from the 3D scans~\cite{pons2017clothcap}. Differentiable simulation techniques~\cite{li2022diffcloth, Liang19_DiffCloth} become requisite for such inverse problems. Prior arts are often designed for inverse problems of a handful of parameters. They either become prohibitive or do not converge when the inverse problem has a large number of unknown variables to be optimized. 
Deep neural networks offer a powerful modality for extracting knowledge from observations. They have also been applied for garment and cloth animation. For instance, \citet{santesteban2019learning} show that deep nets can be exploited to predict garment deformation based on physics-based simulation results.  PBNS~\cite{bertiche2020pbns} and SNUG~\cite{santesteban2022snug} use unsupervised learning for the synthesis of garment deformation. \citet{bertiche2022neural} further generalizes this idea to learn dynamic garment movements. However, existing methods mostly focus on sheet-based models. A learning-based yarn-level model remains under-explored. This is likely due to the lack of high-quality training data and efficient simulation algorithms. 


\paragraph{Numerical methods}
An efficient forward simulation is often the key ingredient of the inverse problem. As the bottleneck in cloth simulation is often at solving the energy Hessian, a natural thought is to avoid a full linear solve in classic Newton's method. Following this idea, \citet{hecht2012updated} propose a lagged factorization scheme that reuses existing Cholesky factorization to save
the computation. Multi-resolution~\cite{capell2002multiresolution,lee2010multi} and multigrid~\cite{tamstorf2015smoothed,wang2018parallel} solvers project fine-grid residual errors onto a coarser grid, on which linear or nonlinear iterations are more effective~\cite{bolz2003sparse,tamstorf2015smoothed,xian2019scalable,zhu2010efficient}. \citet{liu2013fast} treat the implicit Euler integration as an energy minimization problem, in which spring constraints can be solved in parallel in the local step, while the global linear system remains constant on run-time. Projective Dynamics (PD)~\cite{bouaziz2023projective} extend this concept to support a wider range of hyperelastic material models in quadratic form. Additionally, the computation of integration has been accelerated using methods such as the Chebyshev~\cite{Wang2015Chebyshev,wang2016Descent}, Gauss-Seidel~\cite{Fratarcangeli2016Vivace}, and various parallelization techniques on GPU~\cite{wang2016Descent,Fratarcangeli2016Vivace,wu2020repulsioncloth}. \highlight{Fully leveraging the obtained gradient to find a better search direction to improve convergence is also essential for achieving better performance. A common strategy involves estimating the second-order information of the system, e.g. L-BFGS\cite{du2021diffpd,li2022diffcloth} or Anderson acceleration\cite{Peng2018anderson}. \citet{Wang2018rulefree} and \citet{Zimmermann2019puppet} propose using the Gauss-Newton solver, which exhibits promising convergence. However, assembling and factorizing the dense Gauss-Newton solver becomes a new bottleneck. Similar to how the adjoint method circumvents dense matrices, \citet{Zehnder2021} introduces two additional adjoint variables that transform the Gauss-Newton solver to be a sparse linear system, greatly accelerating the performance.} 

\begin{figure*}[ht!]
  \centering
  \includegraphics[width=\linewidth]{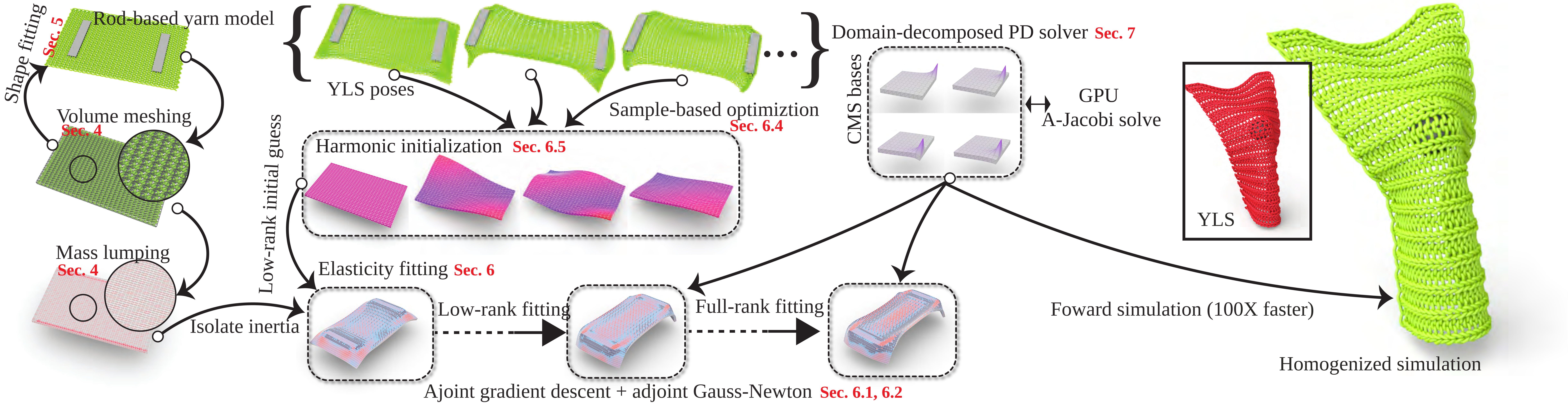}
  \caption{\textbf{\highlight{Volumetric homogenization} pipeline.}~~Given an input rod-based yarn model, our pipeline generates a volume mesh encapsulating the entire yarn structure. We lump the yarn-level mass to the mesh's nodal points so that the inertia effect(\autoref{eq:g}) can be isolated. The yarn-to-mesh and mesh-to-yarn shape fitting facilitates us to define the loss function for homogenization. Per-element material parameters are obtained via an adjoint Gauss-Newton procedure, in a sample-by-sample manner. With a domain-decomposed PD solver, our method \highlight{not only} produces high-quality animation of knitwear garments that is visually similar to the full-scale YLS result \highlight{but also achieves these results two orders of magnitude faster.}}
  \label{fig:pipeline}
  \Description{}
\end{figure*}

\paragraph{Homogenization \& coarsening}
Our work is also closely related to computational homogenization~\cite{andreassen2014determine,allaire2005multiscale,grinspun2002charms} and numerical coarsening~\cite{chen2017dynamics,chen2018numerical}. This category of computing techniques aims to extract a low-rank representation of complex systems via global (coarse mesh) to local (fine mesh) optimization so that simulation at runtime efficiently captures the correct dynamics even on coarse grids~\cite{kharevych2009numerical,nesme2009preserving}. It has been successfully applied for fabrication~\cite{panetta2015elastic,chen2017dynamics} or for accelerated simulation~\cite{torres2016high}. \citet{chen2018numerical} fit a material-aware shape function so that the coarse mesh replicates the behavior of a fine model. Constitutive model homogenization focuses more on \highlight{macroscopic} material responses~\cite{de2015rve,blanco2016variational}, where the underlying strain-stress relation is synthesized at \highlight{RVE} (representative volume element). This method has also been applied to the inverse problem of micro-structure optimization or topology optimization~\cite{eschenauer2001topology}. \highlight{For instance, \citet{schumacher2015microstructures} build deformable objects with target stiffness using cells of prescribed bulking behavior.} \citet{chen2015data} propose a data-driven extension of this potential energy fitting idea to non-linear materials where now a coarse constitutive
model is found through linear regression based on a set of deformation samples obtained from random forcing.

Our method shares many similarities in algorithmic rationale and method design with those prior arts and specifically targeting realistic and efficient knitwear simulation. Our method is directly relevant and strongly inspired by recent contributions 
on HYLC \cite{sperl2020homogenized,Sperl2022yarnestimate}. 
HYLC shows a sheet-level homogenization paradigm that regresses a spline-based constitutive model. The homogenization substantially reduces the DOF, and the sheet-level simulation is much faster than YLS. The major difference between our method and HYLC is that our homogenization escalates the dimensionality of coarse mesh i.e., from a sheet- or rod-based discretization to a volumetric one. On the surface, such a strategy does not align with the original motivation of homogenization or coarsening, as it uses more DOF. Yet, we show that the increased dimensionality simplifies the material design 
when incorporated with a
volume preservation constraint to homogenize bending and twisting effects of codimensional models. We are not the first to leverage volume preserving to model nonlinear bending deformation. \citet{chen2023multi} use a volumetric prism element to simulate thick garments and have demonstrated the feasibility of this approach. Another major difference is that we learn a set of independent material parameters at each volume element, resulting in a high-dimensional space-time optimization problem. We name our method \emph{\highlight{volumetric} homogenization}, hinting at these novel features of our formulation. The complexity of \highlight{volumetric homogenization} is unlike most prior methods. To address this, we propose to leverage a domain-decomposed PD solver and \highlight{employ} adjoint Gauss-Newton\highlight{\cite{Zehnder2021}} to make the homogenization manageable. \highlight{Volumetric} homogenization is more expressive than HYLC--we can replicate the dynamics of complex and non-repetitious knits. It is also more efficient--the use of simpler material models with a fast solver \highlight{offsets} the increase in homogenized DOF.

\section{Method Overview}

Our framework allows stable and efficient knitwear simulation that 
closely mimics the behavior
of a full-scale YLS. To achieve this objective, our pipeline estimates a spatially varying homogenization scheme for the yarn fabric. Unlike prior methods that homogenize a triangular mesh approximating the fabric mid-surface~\cite{sperl2020homogenized,fengneural2024} and assign uniform material parameters across the mesh, our pipeline, as shown in \autoref{fig:pipeline}, begins by constructing a volumetric mesh to encapsulate the given yarn-level model and computes the mass for each nodal point (\autoref{sec:construction}). With the sequence of yarn-level simulated results, we proceed to compute the best-fitting material parameter for each volume element. This is a difficult task, because we do not have bounding volumetric mesh for each frame, the material parameters over mesh are of high dimension, and the optimization is sensitive to the initial condition. Therefore, we decompose the problem into two fitting steps. First, we formulate an optimization problem to find the best-fitting mesh shape matching the yarn-level deformation (\autoref{sec:shapefitting}). With the fitted shape, we employ an adjoint Gauss-Newton formulation\highlight{\cite{Zehnder2021}} that estimates a quasi-second-order descent direction of the high-dimension material parameter (\autoref{sec:elasticity_fitting}). Adjoint Gauss-Newton synergizes with a harmonic initialization algorithm to progressively explore a good initial guess. To utilize the homogenized mesh generated from our fitting pipeline, we introduce a novel GPU-based forward simulator (\autoref{sec:dd_pd}). The global PD matrix is partitioned into domains corresponding to different knit patterns, and we leverage component mode synthesis (CMS)~\cite{craig1988block} to build a subspace preconditioner for the global stage solve, followed by a full-space GPU-based Jacobi iteration.

\section{Mesh construction} \label{sec:construction}
Given a yarn-level model discretized into piece-wise line segments at the rest pose, we construct a volumetric mesh to encapsulate the yarn structure and assign the mass for each nodal point so it can facilitate the subsequent fitting steps and mesh-level simulation.
\setlength{\columnsep}{5 pt}
\begin{wrapfigure}{r}{0.55\linewidth}
   \includegraphics[width=\linewidth]{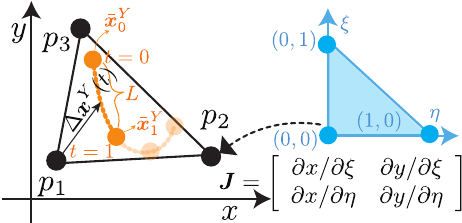}
   \caption{\textbf{Mass lumping.}~~We estimate lumped mass at each mesh node via the line integral of the shape function over the embedded yarn segment.}\label{fig:mass_lumping}
   \Description{}
\end{wrapfigure}
In mesh construction, we employ the 3D Bresenham algorithm~\cite{vzalik1997efficient,zhang2018efficient} to create a volumetric grid, such that each voxel encapsulates a non-zero portion of a yarn thread. To ensure connectivity and eliminate dangling voxels, additional voxels are included in rare cases where the thread passes through a voxel corner. To prevent numerical locking of the hexahedral element~\cite{bathe2006finite}, each voxel is further split into six tetrahedrons. Let $n^Y$ and $n^{\highlight{V}}$ be the total DOF for the yarns and the \highlight{volume} mesh, respectively. Typically, $n^{\highlight{V}}$ is around half of $n^Y$. In other words, meshing the yarn model is unable to lower the DOF count by orders like in~\citet{sperl2020homogenized}. As we discuss later, our method excels in a more stable energy formulation and highly parallelizable GPU solver after the volumetric homogenization. This overcomes the cost of high DOF and makes the entire pipeline much more efficient than prior methods~\cite{fengneural2024,sperl2020homogenized}.



The next step is to distribute the yarn curves to the surrounding nodal points of the mesh. A 2D illustration is shown in \autoref{fig:mass_lumping}, where the triangle element encloses a piece of yarn thread of three linearized segments. Let $\bm{x}$ and $m$ be the position and mass of a material particle in the element. It is assumed that
\begin{equation}
m = \sum_i N_i(\xi, \eta) \cdot m_i \;,    
\end{equation}
where $m_i$ represents the lumped mass of the $i$-th nodal point. $N_i(\xi, \eta)$ is the shape function defined with the natural coordinates
\begin{align*}
    N_1(\xi, \eta) &= 1 - \xi - \eta\;, &
    N_2(\xi, \eta) &= \xi\;, &
    N_3(\xi, \eta) = \eta \;. 
\end{align*}
We compute the lumped mass for each nodal point as
\begin{equation}\label{eq:mass_fit}
    m_i = \sum_j L_j\int_0^1 N_i \left(\bm{J}^{-1} \Delta \bm{x}_j^Y(t)\right) \rho_j(t) \mathrm{d} t \;,
\end{equation}
where the summation index $j$ iterates over all linearized yarn segments (three segments in this example), $L_j$ is the length of the $j$-th segment, $\bm{J}$ is the element Jacobi, which relates the natural coordinates with the material coordinates, and $\Delta \bm{x}_j^Y(t)$ is the vector in the material space from the natural origin to the parameterized yarn-level position on $j$-th segment. For instance, the first segment in \autoref{fig:mass_lumping} has two endpoints $\bar{\bm{x}}_0^Y$ and $\bar{\bm{x}}_1^Y$, and we have
\begin{equation}
    \Delta \bm{x}^Y(t) = \bar{\bm{x}}_0^Y + t\left(\bar{\bm{x}}_1^Y - \bar{\bm{x}}_0^Y\right) - \bar{\bm{x}}_0\;.
\end{equation}
Note that we use the overbar $\bar{(\cdot)}$ to denote the rest-shape position.

\section{Shape fitting} \label{sec:shapefitting}

With the mesh constructed at the rest pose, our next step is to estimate its shape based on the simulated yarn-level data for each frame. 
This boils down to finding out the mesh position given the displacement of embedded yarn threads. We use superscripts $(\cdot)^Y$ and $(\cdot)^{\highlight{V}}$ to differentiate variables defined on yarns or on the \highlight{volume} mesh. Let $\bm{x}^{\highlight{V}} \in \mathbb{R}^{3 n^{\highlight{V}}}$ be mesh-level position, stacking $x$, $y$, and $z$ coordinates of all the nodes on the tetrahedral mesh. The mesh-to-yarn (\highlight{V}2Y) transfer $\phi^{{\highlight{V}}2Y}: \mathbb{R}^{3 n^{\highlight{V}}} \rightarrow \mathbb{R}^{3 n^Y}$ is conveniently established with shape-function-based interpolation
\begin{equation}\label{eq:m2y}
\phi^{{\highlight{V}}2Y}(\bm{x}^{\highlight{V}}) \triangleq \bm{N}\bm{x}^{\highlight{V}} \;,    
\end{equation}
where $\bm{N} \in \mathbb{R}^{3 n^{\highlight{V}} \times 3 n^Y}$ contains $N_i$ values at yarn endpoints. However, the map along the other direction i.e., from yarn to mesh or Y2\highlight{V}, is less straightforward.

We formulate the Y2\highlight{V} transfer $\phi^{Y2\highlight{V}}$ by minimizing position and deformation inconsistency given an input yarn pose $\bm{x}^Y\in \mathbb{R}^{3 n^Y}$. 
The deformation error measures the discrepancy of the deformation gradient tensor between yarn and the encapsulating volumetric mesh. The deformation gradient of a yarn segment is computed as:
\begin{align}
\bm{F}^Y (\bm{x}^Y) &= \left[\bm{x}^Y_1 - \bm{x}^Y_0, \bm{n}^Y_1, \bm{n}^Y_2\right]  \left[\bar{\bm{x}}_1^Y - \bar{\bm{x}}_0^Y, \bar{\bm{n}}^Y_1, \bar{\bm{n}}^Y_2\right]^{-1} \nonumber 
\end{align}
where $\bar{\bm{x}}_{0,1}^Y$ and $\bm{x}_{0,1}^Y$ are the rest-shape and deformed positions of two endpoints and $\bar{\bm{n}}_{0,1}^Y$ and $\bm{n}_{0,1}^Y$ are two mutually perpendicular material normals. \highlight{The deformation gradient includes a rotation and stretch component. While the stretch component can be directly averaged to the corresponding element, the rotation component cannot.}  \highlight{To manage the averaging of the rotation component, we apply polar decomposition to $\bm{F}^Y$, which yields a rotation tensor and a symmetric deformation tensor such that $\bm{F}^Y = \bm{R}^Y \bm{S}^Y$. We parameterize the rotation tensor using matrix exponential as $\bm{R}^Y = \mathsf{exp}(\bm{\Omega}^Y)$. Note that $\bm{\Omega}^Y$ is a skew-symmetric matrix, which can be directly averaged using weighted summation. This geometrically corresponds to averaging the rotation axis and rotation angle\cite{matexp1998}.} Based on this information, we estimate the deformation gradient of the element as
\begin{equation}
    \bm{F}^{Y2\highlight{V}} (\bm{x}^Y)= \mathsf{exp}\left(\frac{1}{\sum_j L_j} \sum_j L_j \bm{\Omega}_j^Y \right) \left(\frac{1}{\sum_j L_j}\sum_j L_j \bm{S}^Y_j\right)\;.
\end{equation}
As in \autoref{eq:mass_fit}, the summation is for all the yarn segments in the element.

We define the Y2\highlight{V} transfer as the minimizer of the optimization
\begin{multline}\label{eq:y2m}
\phi^{Y2\highlight{V}}(\bm{x}^Y) \triangleq
\arg\min_x \sum_e V_e \left\| \mathcal{D}\cdot(\bm{B}_e\bm{x}) - \bm{F}^{Y2\highlight{V}}_e\right\|^2_F \\
+ \alpha \left\|\bm{N}\bm{M}^{\highlight{V}}\bm{x} - \bm{M}^Y\bm{x}^Y\right\|^2.    
\end{multline}
Here, $\bm{B}_e \in \mathbb{R}^{12 \times 3 n^Y}$ is a binary matrix picking 12 DOFs for the tetrahedron $e$ from $\bm{x}$ i.e., $\bm{x}_e = \bm{B}_e \bm{x}$. $\mathcal{D} \in\mathbb{R}^{3 \times 3 \times 12}$ is a third-order tensor. It works as a differential operator to compute the deformation gradient of the element out of $\bm{x}_e$. $\|\cdot\|_F$ denotes the matrix Frobenius norm. $\alpha$ is a hyperparameter, and we set it as $\alpha = 0.1$. $\bm{M}^{\highlight{V}}$ is the diagonal lumped mass matrix of the mesh (i.e., using Eq.~\eqref{eq:mass_fit}), and $\bm{M}^Y$ is the mass matrix for the yarn. $V_e$ is the volume of the element. In other words, Y2\highlight{V} transfer is essentially a volumetric Possion reconstruction~\cite{kazhdan2006poisson}--we expect $\phi^{Y2\highlight{V}}(\bm{x}^Y)$ to capture the yarn-level deformation while using $\left\|\bm{N}\bm{M}^{\highlight{V}}\bm{x} - \bm{M}^Y\bm{x}^Y\right\|^2$ as a mass-weighted regularization penalty. 


Shape fitting retrieves the information of the mesh from YLS and allows us to estimate inertia forces \highlight{as defined in \autoref{eq:g}}, derived from an arbitrary yarn motion sequence. As a result, the yarn-level deformation can be understood as a quasi-static one under the non-inertia frame, and the \highlight{volumetric homogenization} only needs to focus on fitting the elastic material since mass is decoupled.

\section{Elasticity fitting}\label{sec:elasticity_fitting}
Elasticity fitting lies at the core of our \highlight{volumetric homogenization} pipeline. In a nutshell, elasticity fitting estimates spatially varying material parameters for each element given input YLS sequences. This is particularly challenging because 1) \highlight{volumetric homogenization} seeks a high-dimension material parameter with a large number of \highlight{unknowns}, and 2) the optimization takes consideration of all yarn poses. \highlight{We employ several techniques to address these challenges including the use of second-order Gauss-newton, as described by \citet{Zehnder2021}, within the framework of the adjoint method. Additionally, we introduce a novel harmonic initialization scheme for initial value guessing.} We borrow the idea of stochastic descent~\cite{bottou1991stochastic} to mitigate the dimensionality concern of space-time optimization. 

Let $n^E$ be the total number of elements on the mesh. The material vector $\bm{\gamma}$ has ${2 n^E}$ freedoms such that an element $e$ has two hypothesized material parameters, namely $\gamma_e^s$ and $\gamma_e^v$. We define an intuitive elastic energy at each element $e$ as
\begin{equation}\label{eq:energy}
    E_e = \gamma_e^s V_e\left\| \bm{F}_e - \bm{R}(\bm{F}_e)\right\|^2_F + \gamma_e^v V_e \left\| \bm{F}_e - \bm{V}(\bm{F}_e)\right\|^2_F,
\end{equation}
where
\begin{equation}\label{eq:pd_local}
    \bm{R} = \arg\min_{\bm{A} \in \mathsf{SO}(3)} \| \bm{F} - \bm{A}\|^2_F\quad\text{and}\quad \bm{V} = \arg\min_{\bm{A} \in \mathsf{SL}(3)} \| \bm{F} - \bm{A}\|^2_F.
\end{equation}
Here, $\mathsf{SL}(3)$ is the special linear group that preserves the volume (i.e. $|\bm{V}| = 1$), $\mathsf{SO}(3)$ is the special orthogonal group that preserves the length, $\gamma_e^s$ determines the strength of penalizing the strain magnitude, and $\gamma_e^v$ gives the strength of preserving the volume. 

\paragraph{The choice of energy} There exists a wide range of choices for energy formulation, and many commonly seen hyperelastic energies should serve the purpose well. Our primary argument is that the material complexity holds less significance compared to the material variation, which naturally reflects the spatial adaptivity inherent in knit patterns. The elastic model of \autoref{eq:energy} is simple--its constraint-based quadratic form eases the implementation efforts and produces good results in practice. 


Elasticity fitting finds the optimal $\bm{\gamma}$ i.e., the value of $\gamma_e^s$ and $\gamma_e^v$ for all $n^E$ elements, such that \highlight{V}2Y transfer (\autoref{eq:m2y}) of the resultant mesh simulation, $\bm{x}^{\highlight{V}} (\bm{\gamma})$, constitutes a similar yarn dynamics obtained from the full-scale YLS. For a YLS sequence of $n^F$ frames
\begin{equation*}
    \left\{ \bm{x}^Y_1, \bm{x}^Y_2, \bm{x}^Y_3, \cdots, \bm{x}^Y_{n^F}\right\},
\end{equation*}
we build an error metric or the loss function between $\bm{x}^Y_i$ and $\bm{x}^{\highlight{V}}_i(\bm{\gamma})$ for $i = 1,\cdots,n^F$ and minimize the accumulated error for the entire sequence
\begin{equation}\label{eq:elasticity_fit}
    \arg\min_{\bm{\gamma}} \sum_{i=1}^{n^F}\epsilon_i(\bm{\gamma}),\quad\text{s.t.}\;\bm{\gamma} \geq 0,
\end{equation}
where
\begin{multline}\label{eq:loss}
    \epsilon_i(\bm{\gamma}) = \sum_e V_e \left\| \mathcal{D}\cdot\left(\bm{B}_e\bm{x}^{\highlight{V}}_i(\bm{\gamma})\right) - \bm{F}^{Y2{\highlight{V}}}_e(\bm{x}^Y_i)\right\|^2_F \\
+ \alpha \left\|\bm{N}\bm{M}^{\highlight{V}}\bm{x}^{\highlight{V}}_i(\bm{\gamma}) - \bm{M}^Y\bm{x}^Y_i\right\|^2.    
\end{multline}
Here, $\epsilon_i$ adopts the same error measure as in Y2{\highlight{V}} transfer (\autoref{eq:y2m}) but it now depends on $\bm{\gamma}$ since $\bm{x}^{\highlight{V}}_i(\bm{\gamma})$ relates to $\bm{\gamma}$ by the mesh-level simulation. The sub-index $i$ here refers to the index of the input $n^F$ YLS poses.

\autoref{eq:elasticity_fit} describes a space-time optimization problem. It imposes a significant computational challenge since both $n^F$ and $n^E$ are big numbers, not to mention $n^{\highlight{V}}$ and $n^Y$ are also of high resolution. While first-order descent methods like gradient descent are often preferred due to their simplicity, they fail to deliver a good result even after a large number of iterations in this case. We have to resort to more sophisticated optimization techniques of a higher order for elasticity fitting of \autoref{eq:elasticity_fit}.

\subsection{Adjoint Gradient Descent}
Let us start with the gradient. Expanding the gradient of $\epsilon_i$ via the chain rule yields
\begin{equation}
\frac{\partial \epsilon_i}{\partial \bm{\gamma}} = \frac{\partial \epsilon_i}{\partial \bm{x}^{\highlight{V}}_i}  \cdot \frac{\partial \bm{x}^{\highlight{V}}_i}{\partial \bm{\gamma}}.
\end{equation}
The specific form of ${\partial \bm{x}^{\highlight{V}}}/{\partial \bm{\gamma}}$ is up to the mesh-level simulation result, which is often formulated as a variational optimization minimizing the sum of the inertial potential $I$ and the elasticity potential $\sum E_e$
\begin{align}\label{eq:var}
    &\arg \min_{\bm{x}^{\highlight{V}}_i} I + \sum_e E_e, 
    &&\text{where}&
    I &=\frac{1}{2 \Delta t^2} \bm{a}^{{\highlight{V}}^\top}_i \bm{M}^{{\highlight{V}}} \bm{a}^{\highlight{V}}_i.
\end{align}
Shape fitting allows us to approximate $\bm{a}_i^{\highlight{V}}$ via
\begin{equation}\label{eq:app_a}
\bm{a}_i^{\highlight{V}} \approx \phi^{Y2{\highlight{V}}}\left(\bm{x}_i^Y - 2\bm{x}^Y_{i-1} + \bm{x}^Y_{i-2}\right) - \Delta t^2 \bm{M}^{{\highlight{V}}^{-1}}\bm{f}^{\highlight{V}}_i,    
\end{equation}
so that it becomes a known vector and depends on $\bm{x}^Y_i$, $\bm{x}^Y_{i-1}$, $\bm{x}^Y_{i-2}$, and the external force $\bm{f}^{\highlight{V}}_i$. In other words, the Y2\highlight{V} transfer function $\phi^{Y2{\highlight{V}}}$ and pre-computed nodal mass (\autoref{eq:mass_fit}) decouple $I$ from the simulation, converting a dynamic problem to a quasi-static one.

The necessary optimality condition of \autoref{eq:var} gives
\begin{equation}\label{eq:g}
    \bm{g}(\bm{x}^{\highlight{V}}_i, \bm{\gamma}) \triangleq \underbrace{\bm{M}^{\highlight{V}} \phi^{Y2{\highlight{V}}} \left( \bm{x}_i^{\highlight{Y}} - 2\bm{x}^Y_{i-1} + \bm{x}^Y_{i-2} \right) - \bm{f}_i^{\highlight{V}}}_{\text{inertia force + external force}} + \frac{\partial \sum E_e}{\partial \bm{x}_i^{\highlight{V}}} = 0.
\end{equation}
It is easy to see that \autoref{eq:g} is the requirement of force equilibrium, which is the constraint that should always be satisfied, making it an identity equation. Differentiating \autoref{eq:g} at both sides w.r.t. $\bm{\gamma}$ yields
\begin{equation}\label{eq:pxpgamma}
    \frac{\mathrm{d} \bm{g}}{\mathrm{d} \bm{\gamma}} = \frac{\partial \bm{g}}{\partial \bm{x}^{\highlight{V}}_i} \cdot \frac{\partial \bm{x}^{\highlight{V}}_i}{\partial \bm{\gamma}} + \frac{\partial \bm{g}}{\partial \bm{\gamma}} = 0 \Rightarrow \frac{\partial \bm{x}^{\highlight{V}}_i}{\partial \bm{\gamma}} = -\left(\frac{\partial \bm{g}}{\partial \bm{x}^{\highlight{V}}_i}\right)^{-1} \cdot \frac{\partial \bm{g}}{\partial \bm{\gamma}}.
\end{equation}
To avoid solving the linear system of $\partial \bm{g}/\partial \bm{x}^{\highlight{V}}_i$ for $2 n^E$ times (recalling $\bm{\gamma}$ is a $2 n^E$-dimension material vector and, therefore, ${\partial \bm{g}/\partial \bm{x}^{\highlight{V}}_i \in \mathbb{R}^{3n^{\highlight{V}} \times 2n^E}}$), the adjoint method \cite{tarantola2005inverse,givoli2021tutorial} leverages an adjoint state vector $\bm{\lambda}$, and evaluates the gradient of the target loss function via
\begin{equation}\label{eq:adjoint_state}
    \frac{\partial \epsilon_i}{\partial \bm{\gamma}} = -\bm{\lambda}^\top \frac{\partial \bm{g}}{\partial \bm{\gamma}},
\end{equation}
where $\bm{\lambda}$ is obtained by solving the linear system of
\begin{equation}\label{eq:adjoint_state_solve}
    \left(\frac{\partial \bm{g}}{\partial \bm{x}^{\highlight{V}}_i}\right) \bm{\lambda} = \left(\frac{\partial \epsilon_i}{\partial \bm{x}^{\highlight{V}}_i}\right)^\top.
\end{equation}
The gradient of the loss function $\partial \epsilon_i / \partial \bm{\gamma}$ allows us to employ first-order optimizers, e.g., gradient descent, to iterative refine the material vector. Unfortunately, it is known that first-order methods are less effective as $\bm{\gamma}$ approaches a local optimum. To improve the convergence, we employ a hybrid optimization scheme, which uses different optimizers at different stages of the elasticity fitting process.

\subsection{Adjoint Gauss-Newton}
A well-known strategy to improve the convergence of gradient descent is to regularize the gradient by a preconditioning matrix $\bm{P}_i$, often an SPD (symmetric positive definite) matrix. For instance, Newton's method uses the inverse of the Hessian matrix
\begin{equation}\label{eq:hessian}
    \bm{H}_i = \frac{\partial^2 \epsilon}{\partial \bm{\gamma}^2} = \left(\frac{\partial \bm{x}^{\highlight{V}}_i}{\partial \bm{\gamma}}\right)^\top \cdot \frac{\partial^2 \epsilon_i}{\partial \bm{x}^{{\highlight{V}}^2}_i} \cdot \frac{\partial \bm{x}^{\highlight{V}}_i}{\partial \bm{\gamma}} + \frac{\partial \epsilon_i}{\partial \bm{x}^{\highlight{V}}_i} \cdot \frac{\partial^2 \bm{x}_i^{\highlight{V}}}{\partial \bm{\gamma}^2},
\end{equation}
so that the preconditioned search direction becomes $\bm{H}^{-1}_i ({\partial \epsilon_i}/{\partial \bm{\gamma}})^\top$. When $\|\partial \epsilon_i / \partial \bm{\gamma}\|$ is small, Newton's method delivers locally second-order convergence~\cite{nocedal1999numerical}.

We observe that when $\epsilon_i$ is small and mesh deformation $\bm{x}^{\highlight{V}}_i$ starts aligning with the target yarn pose $\bm{x}^Y$, $\partial \epsilon_i / \partial \bm{x}^{\highlight{V}}_i$ is always close to zero. This allows an alternative precondition matrix $\bm{P}_i$ that discards the second term of $\mathbf{H}_i$, such that
\begin{equation}
    \bm{P}_i = \left(\frac{\partial \bm{x}^{\highlight{V}}_i}{\partial \bm{\gamma}}\right)^\top \bm{G}_i  \frac{\partial \bm{x}^{\highlight{V}}_i}{\partial \bm{\gamma}}\;,
\end{equation}
where, $\bm{G}_i = {\partial^2 \epsilon_i}/{\partial \bm{x}^{{\highlight{V}}^2}_i}\in\mathbb{R}^{3n^{\highlight{V}} \times 3n^{\highlight{V}}}$ and can be pre-computed. This Hessian simplification strategy is a.k.a. Gauss-Newton method, and the preconditioned search direction can be obtained by solving the linear system $\bm{P}_i$:
\begin{equation}\label{eq:gn}
    \bm{P}_i \bm{d}_i^{GN} = -\left(\frac{\partial \epsilon_i}{\partial \bm{\gamma}}\right)^\top,
\end{equation}
where the loss gradient on the r.h.s. of \autoref{eq:gn} can be obtained via the adjoint solve (\autoref{eq:adjoint_state} and~\autoref{eq:adjoint_state_solve}).

While the formulation is well known, Gauss-Newton is seldom used with an adjoint method in practice. The difficulty lies in the fact that $\partial \bm{x}^{\highlight{V}}_i / \partial \bm{\gamma}$ should never be explicitly evaluated via solving \autoref{eq:pxpgamma} and, therefore, we do not have the actual precondition matrix $\bm{P}_i$. Even if we could compute $\bm{P}_i$ exactly, solving \autoref{eq:gn} remains prohibitive since $\bm{P}_i$ is likely a dense matrix.

\highlight{\citet{Zehnder2021} demonstrate that this issue can be circumvented. Similar to how the adjoint method avoids explicit computation of $\frac{\partial x^{M}_{i}}{\partial \gamma}$, two additional adjoint state vectors are introduced to transform the dense system \autoref{eq:gn} into the following larger sparse linear system:}
\begin{equation}\label{eq:agn}
\left[
\begin{array}{ccc}
    \bm{G}_i & - \left(\frac{\partial \bm{g}}{\partial \bm{x}^{\highlight{V}}_i}\right)^\top & \bm{0} \\
    -\frac{\partial \bm{g}}{\partial \bm{x}^{\highlight{V}}_i} & \bm{0} & \frac{\partial \bm{g}}{\partial \bm{\gamma}}\\
    \bm{0} & \left(\frac{\partial \bm{g}}{\partial \bm{\gamma}}\right)^\top & \bm{0}
\end{array}
\right]
\left[
\begin{array}{c}
\bm{\mu}\\
\bm{\nu}\\
\bm{d}_i^{GN}
\end{array}
\right]
=
\left[
\begin{array}{c}
\bm{0}\\
\bm{0}\\
\left(\frac{\partial \epsilon_i}{\partial \bm{\gamma}}\right)^\top
\end{array}
\right].
\end{equation}
One can quickly verify that the solution of \autoref{eq:agn} coincides with the solution of \autoref{eq:gn}: from the first two lines of \autoref{eq:agn}, we have $(\partial \bm{g} / \partial \bm{x}^{\highlight{V}}_i)^{-\top}\bm{G}_i \bm{\mu} = \bm{\nu}$ and $(\partial \bm{g} / \partial \bm{x}^{\highlight{V}}_i)^{-1} (\partial \bm{g} / \partial \bm{\gamma}) \bm{d}_i^{GN} = \bm{\mu}$, suggesting
\begin{equation*}
    \left(\frac{\partial \bm{g}}{\partial \bm{x}_i^{\highlight{V}}}\right)^{-\top} \bm{G}_i \underbrace{ \left(\frac{\partial \bm{g}}{\partial \bm{x}_i^{\highlight{V}}}\right)^{-1} \frac{\partial \bm{g}}{\partial \bm{\gamma}}}_{-\partial \bm{x}_i^{\highlight{V}} / \partial \bm{\gamma} } \bm{d}_i^{GN} = \bm{\nu}.
\end{equation*}
Left multiplying $(\partial \bm{g} / \partial \bm{\gamma})^\top$ both sides and knowing \highlight{$(\partial \bm{g} / \partial \bm{\gamma} )^\top \bm{\nu} = \partial \epsilon_i / \partial \bm{x}_i^{\highlight{V}} $}, i.e. the third row of \autoref{eq:agn}, restores the formula back to the vanilla Gauss-Newton of \autoref{eq:gn}. 

When $\bm{\gamma}$ is fixed, $\epsilon_i$ becomes a quadratic form of $\bm{x}^{\highlight{V}}_i$, and $\bm{G}_i$ is therefore non-negative definite. To this end, we follow the Levenberg-Marquardt method~\cite{more2006levenberg} that adds a small diagonal $\kappa \bm{I}$ at l.h.s. of \autoref{eq:agn} to secure its positive definiteness. It is noteworthy that $\partial \bm{g} / \partial \bm{x}_i^{\highlight{V}}$ is a sparse matrix--it has non-zero entities only at the adjacent mesh elements (similar to FEM matrices), and $\partial \bm{g} / \partial \bm{\gamma}$ is also a sparse matrix (only the element's material parameter influences its residual force). Solving \autoref{eq:agn} is more efficient than solving \autoref{eq:gn} despite the increased matrix dimension.

\subsection{Linear Search \& Non-Negativity Constraint}

Our optimization is two-stage. We use the vanilla gradient descent based on the adjoint method (\autoref{eq:adjoint_state_solve}) for the first batch of iterations (10 - 15 iterations in our implementation). Gradient descent remains a competitive option at the early stage of the optimization due to its efficiency. Adjoint Gauss-Newton (\autoref{eq:agn}) then ensues, which offers a stronger descent direction. We observe that 20 - 30 Gauss-Newton iterations are more effective than over 1,000 gradient descent in the later phase of the optimization. 

It should be noted that all the components i.e., $\partial \bm{g} / \partial \bm{x}^{\highlight{V}}_i$, $\partial \bm{g} / \partial \bm{\gamma}$, and $\partial \epsilon_i / \partial \bm{\gamma}$ that assemble the Gauss-Newton system are readily available after the gradient computation. Compared with gradient descent, the additional cost of our adjoint Gauss-Newton comes from the linear solve of \autoref{eq:agn}. Thanks to the sparsity of the matrix, this step is \emph{not} the bottleneck of the pipeline. Instead, the most expensive computation is always at finding $\bm{x}^{\highlight{V}}_i$ to satisfy the equilibrium condition i.e., \autoref{eq:g}. As to be discussed in \autoref{sec:dd_pd}, the constraint-based energy formulation (\autoref{eq:energy}) allows the use of highly efficient parallel GPU procedures to accelerate this step, which is not only helpful for the forward simulation but also for the elasticity fitting.

A line search is needed for both gradient descent and Gauss-Newton phases to prevent overshooting. We use the default step size of $0.01$ for gradient descent refinements and $1.0$ for Gauss-Newton refinements. We shrink the step size by half if the refinement produces a higher loss value. 

$\bm{\gamma}$ should be non-negative, and simulation becomes ill-defined if some components $[\gamma]_j$ in $\bm{\gamma}$ are smaller than zero. We use a mixed projection-pivoting strategy to enforce the non-negativity of $\bm{\gamma}$. Specifically, when a refinement $\bm{\gamma} \leftarrow \bm{\gamma} + \Delta \bm{\gamma}$ produces negative components ($[\gamma]_j < 0$), we correct their values to be a small positive quality of $1e-3$ and cache indices of those components/elements. If the next $\Delta \bm{\gamma}$ tries to further reduce the value of those parameters, we clamp them to zero ($[\Delta \gamma]_j \leftarrow 0$) to cancel such constraint-violating refinement. This is an easy but heuristic mechanism to enforce $\bm{\gamma} > \bm{0}$. Occasionally, the clamped search direction becomes non-descent even using adjoint Gauss-Newton. When this occurs, we opt for the pivoting method~\cite{baraff1994fast} that sets most negative $[\gamma]_j$ as equality-constrained ones (i.e., $[\gamma]_j = 1e-3$) and precludes them from the Gauss-Newton solve by removing corresponding rows and columns in the l.h.s. of \autoref{eq:agn}. When pivoting is activated, we do not relax those equality-constrained DOFs. Fortunately, pivoting is rarely needed provided a reasonable initial guess of $\bm{\gamma}$.

\subsection{Sample-by-Sample Fitting}
The discussion so far has focused on elasticity fitting for one input yarn pose $\bm{x}^Y_i$. Recall that the global loss function accumulates across all $n^F$ poses (i.e., see \autoref{eq:elasticity_fit}). This leads to a very high-dimension space-time optimization problem, and solving it in its original form is infeasible.
To ease the computational cost, we optimize $\bm{\gamma}$ sequentially in a frame-by-frame manner. Doing so is similar to stochastic optimization widely used in deep learning~\cite{kingma2014adam}. Specifically, we start with solving the sub-problem of \autoref{eq:elasticity_fit} for the first pose $\bm{x}^Y_i$ as: $\bm{\gamma}_1 \leftarrow \arg\min_{\bm{\gamma}} \epsilon_1(\bm{\gamma}, \bm{x}^Y_1)$. The resulting $\bm{\gamma}_1$ is saved as the current solution of the material vector such that  $\bm{\gamma} \leftarrow \bm{\gamma}_1$. Once $\bm{\gamma}_k$ is computed for the $k$-th pose, we update $\bm{\gamma}$ via by convexly averaging $\bm{\gamma}$ and $\bm{\gamma}_k$ as
\begin{equation}
    \bm{\gamma} \leftarrow \frac{w}{w + w_k} \bm{\gamma} + \frac{w_k}{w + w_k} \bm{\gamma}_k,
\end{equation}
Here, $w, w_k > 0$ suggest the importance of $\bm{\gamma}$ and $\bm{\gamma}_k$. We find the elasticity energy stored in $\bm{x}^Y_k$ to be a good choice of $w_k$, which is invariant under rigid body movements and always non-negative. The weight of $\bm{\gamma}$ is initially set as zero. As the optimization moves forward, we keep tracking the most deformed yarn pose and use the corresponding elastic energy as $w$. 

Theoretically, such an online optimization strategy takes multiple epochs to converge. However, we note that a single pass over all frames always generates high-quality results in practice. Unlike in deep learning, where training data are considered equally important, most YLS poses are redundant or repeating, e.g. poses close in time often have a similar geometry because the motion trajectory of the fabric often remains smooth. This observation motivates us to further sparse the computation by applying elasticity fitting at a subset of fewer sample poses and finding the overall $\bm{\gamma}$ sample by sample. In practice, elasticity fitting over handful poses yields reasonably good results. 

\subsection{Harmonic Initialization}
Another challenge is the high dimensionality of the material vector $\bm{\gamma}$. By assigning each element two independent freedoms, the material versatility is enhanced. On the downside, it also makes the fitting process sensitive to the initial guess of $\bm{\gamma}$. An unlucky guess easily strands the optimization at local minima. To make our per-element homogenization reliable, our pipeline includes a harmonic initialization scheme, as shown in~\autoref{har_init}.

\begin{figure}
  \centering
  \includegraphics[width=\linewidth]{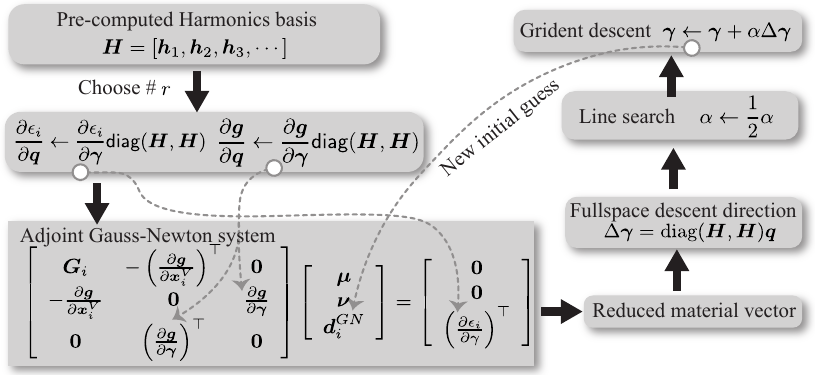}
  \caption{\textbf{Harmonic initialization.}~~We design a progressive initialization strategy based on Harmonic bases of the volume mesh. By projecting the material vector into the Harmonic subspaces of different ranks, \highlight{volumetric homogenization} always finds a reliable initial value for the two-stage elasticity fitting procedure. }\label{har_init}
  \Description{}
\end{figure}

The core idea is progressively exploring the material space from low frequency to high frequency to build the material complexity incrementally. To this end, we split $\bm{\gamma}$ into $\bm{\gamma}^s \in\mathbb{R}^{n^E}$ and $\bm{\gamma}^v \in\mathbb{R}^{n^E}$, which collect $\gamma^s_e$ and $\gamma^v_e$ for all the elements on the mesh respectively. After that, we construct a set of Harmonic bases $\bm{H} = [\bm{h}_1, \bm{h}_2,\cdots, \bm{h}_r]$ by computing $r$ eigenvectors of the mesh Laplacian~\cite{vallet2008spectral,nasikun2018fast} corresponding to the $r$ smallest eigenvalues (see \autoref{fig:harm}). The low-frequency material variation is assumed to be well captured by this set of basis vectors, and we require
\begin{equation}\label{eq:q2gamma}
\bm{\gamma} = 
\left[
\begin{array}{c}
\bm{\gamma}^s\\
\bm{\gamma}^v
\end{array}
\right]
=
\mathsf{diag}(\bm{H}, \bm{H})
\underbrace{
\left[
\begin{array}{c}
\bm{q}^s\\
\bm{q}^v
\end{array}
\right]
}_{\bm{q}} \;,
\end{equation}
where $\bm{q}\in\mathbb{R}^{2r}$ is the generalized material vector of much lower dimension. Since $\bm{H}$ is constant, all the derivatives w.r.t. $\bm{\gamma}$ can be conveniently transferred to $\bm{q}$ by the chain rule:
\begin{equation*}
    \frac{\partial (\cdot)}{\partial \bm{q}} = \frac{\partial (\cdot)}{\partial \bm{\gamma}} \cdot \frac{\partial \bm{\gamma}}{\partial \bm{q}} = \frac{\partial (\cdot)}{\partial \bm{\gamma}} \mathsf{diag}(\bm{H},\bm{H}).
\end{equation*}
For instance, to switch the optimization target to $\bm{q}$ in \autoref{eq:agn}, we right multiply \highlight{$\mathsf{diag}(\bm{H}, \bm{H})$} to $\partial \bm{g} / \partial \bm{\gamma}$ and $(\partial \epsilon_i / \partial \bm{\gamma})^\top$, and sandwich $\bm{G}_i$ with \highlight{$\mathsf{diag}(\bm{H}, \bm{H})$} as
\begin{equation}
\bm{G}_i \leftarrow \mathsf{diag}(\bm{H}^\top, \bm{H}^\top) \bm{G}_i \mathsf{diag}(\bm{H}, \bm{H}).
\end{equation}
The harmonic initialization starts with $r = 1$. In this case, $\bm{H}$ has a single constant-value basis vector $\bm{h}_1$. Projecting $\bm{\gamma}$ into this basis is equivalent to requiring $\bm{\gamma}^s$ and $\bm{\gamma}^v$ to be constant across all the elements. This result is then used as the initial guess for a bigger $r$. The harmonic initialization moves forward progressively with $r = 1$, $r = 10$, and $r = 30$. Afterward, we fit the un-reduced $\bm{\gamma}$ based on the low-rank initialization.

\begin{figure}
  \centering
  \includegraphics[width=\linewidth]{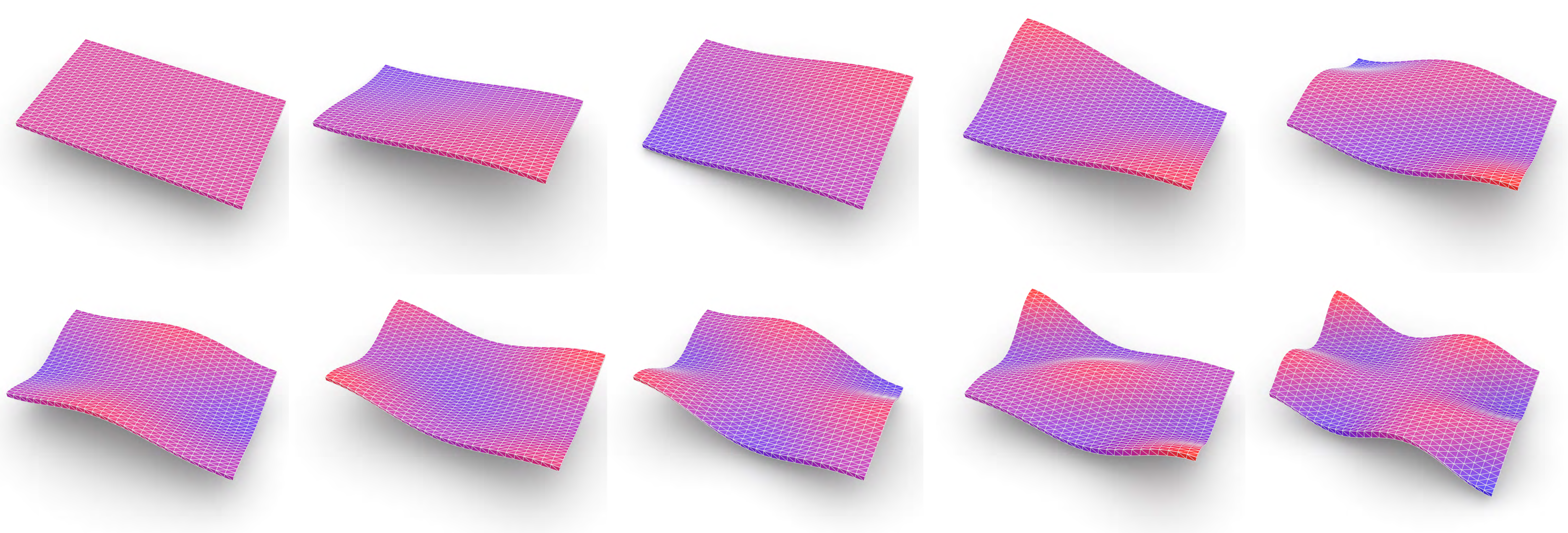}
  \caption{\textbf{Harmonic material bases.}~~We leverage mesh Harmonics to explore a good initial material variation from low frequency to high frequency. This figure visualizes the first ten basis vectors.}\label{fig:harm}
  \Description{}
\end{figure}

\section{Domain-decomposed projective dynamics}\label{sec:dd_pd}
After $\bm{\gamma}$ is obtained, we can simulate the volume mesh at the runtime, which is used to drive the deformation of the underlying yarns via $\phi^{M2Y}$. Our volume embedding implicitly deals with yarn-yarn contacts and collisions. 
This section focuses on the mesh-level simulation, and we omit the superscript $M$ for simplicity of notation. 

As in \autoref{eq:var}, we aim to minimize the total variational energy. Instead of using the approximation in \autoref{eq:app_a}, $\bm{a}$ is now defined as
\begin{equation}
    \bm{a} = \bm{x} - \bm{x}^* - \Delta t \dot{\bm{x}}^*  - \Delta t^2 \bm{M}^{-1} \bm{f},
\end{equation}
where $\bm{x}^*$ and $\dot{\bm{x}}^*$ are the mesh-level position and velocity in the previous time step. Instead of solving unknown position $\bm{x}$ using existing methods, such as Newton's method, we present an efficient domain-decomposed projective dynamics for our constraint-based energy formulation.



\subsection{Projective Dynamics}

Projective dynamics (PD)~\cite{bouaziz2023projective}  splits the variational optimization into global and local steps. At the local step, PD considers the elasticity energy of each element as the shortest square distance to a constraint manifold on which the constraint is exactly satisfied. The key operation is to identify such spot on the constraint manifold i.e., the \emph{target position}. One may notice that our elasticity energy (\autoref{eq:energy} and~\autoref{eq:pd_local}) is formulated exactly in this way. Therefore, the goal of the local step is to find $\bm{R}$ and $\bm{V}$ for each mesh element. The best-fitting rotation can be obtained by polar decomposing the element's deformation gradient $\bm{F}_e$. To obtain the best-fitting volume-preserving transformation $\bm{V}$, we first compute its SVD (singular value decomposition) as $\bm{F}_e = \bm{U}\bm{\Sigma}\bm{W}^\top$. $\bm{\Sigma}$ is a diagonal matrix of three singular values $\sigma_1$, $\sigma_2$, and $\sigma_3$, and the volume-preserving constraint becomes $\sigma_1 \sigma_2 \sigma_3 = 1$. If the element is not inverted or degenerated, $\sigma_1$,  $\sigma_2$, and $\sigma_3$ should be positive. 
We then transfer this constraint into an optimization procedure of three singular values:
\begin{multline}\label{eq:vol_pre}
    \arg\min_{\Delta \sigma_1, \Delta\sigma_2, \Delta\sigma_3} \Delta \sigma_1^2 +  \Delta \sigma_2^2 + \Delta \sigma_3^2,\\
    \text{s.t.}\quad (\sigma_1 + \Delta \sigma_1)(\sigma_2 + \Delta \sigma_2)(\sigma_3 + \Delta \sigma_3) = 1,\\
    \text{and}\quad \sigma_1 + \Delta \sigma_1, \sigma_2 + \Delta \sigma_2, \sigma_3 + \Delta \sigma_3 > 0.
\end{multline}
To solve this nonlinear programming problem, we form its KKT (Karush–Kuhn–Tucker) system only involving the equality constraint using the Lagrange multiplier method. If any of $\Delta \sigma_1$, $\Delta \sigma_2$ or $\Delta \sigma_3$ violates the inequality constraint, we clamp it to $[0.01, +\inf]$ and fix its value at the next iteration. While \autoref{eq:vol_pre} is highly nonlinear, it only has three unknowns. In practice, we always find a good local projection for the volume-preserving constraint in very few iterations.

The global step is a standard linear solve in the form of $\highlight{\bm{K}\bm{x} = \bm{b}}$
where we have
\begin{equation}
    \bm{K} = \left(\frac{\bm{M}}{\Delta t^2}+\sum_e V_e (\gamma^s_e + \gamma^v_e) \bm{B}_e^\top \big(\mathcal{D}^\top : \mathcal{D}\big) \bm{B}_e\right)\bm{x},
\end{equation}
and 
\begin{align}  
    \bm{b} = \frac{\bm{M}}{\Delta t^2} \bm{a} & +  \sum_e V_e\gamma_e^s\bm{B}_e^\top\big(\mathcal{D}^\top : \bm{R}(\bm{B}_e\bm{x}^*)\big) \nonumber \\
    & + \sum_e V_e\gamma_e^v\bm{B}_e^\top\big(\mathcal{D}^\top : \bm{V}(\bm{B}_e\bm{x}^*)\big).
\end{align}
The global step stands as the most expensive computation along the pipeline \highlight{, as the Cholesky decomposition of $K$ may yield a large dense matrix that is not GPU-friendly.} 


\subsection{Domain Decomposition}

\begin{wrapfigure}{r}{0.5\linewidth}
    \includegraphics[width=\linewidth]{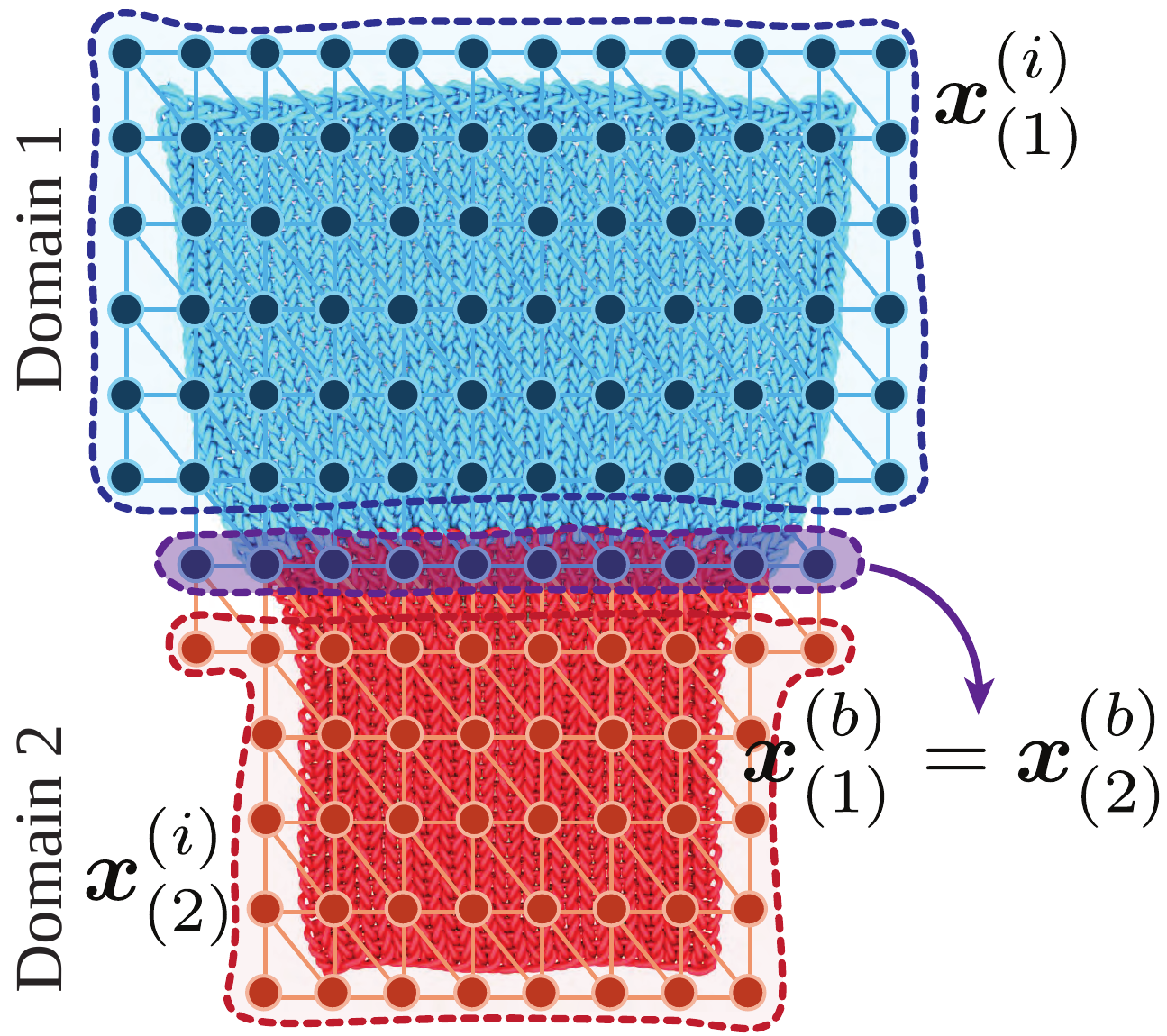}
    \caption{\textbf{DOF types with domain decomposition.}~The fabric is decomposed into two domains corresponding to two different knitting patterns. It is required that duplicated boundary DOFs at domains must always be equal to each other.}
    \Description{}
\end{wrapfigure}
To accelerate the global step computation, we exploit the component mode synthesis (CMS)~\cite{craig1988block,craig1985review} to decompose the system into multiple domains (or components) and build a linear subspace at each domain to analyze the dynamic responses of complex structures. 
CMS was originally designed for linear structural analysis, and its generalization to nonlinear simulation remains an open research problem. Fortunately, the unique modality of PD allows us to apply CMS just at the global stage solve, so we can partition the mesh into multiple domains without aligning the decomposition with the variation of the underlying knit patterns.
Unknown DOFs at each domain can be now grouped into internal DOFs and boundary DOFs. As the name suggests, the boundary DOFs interface with the neighbor domains while the internal DOFs are isolated by the boundary DOFs.

\begin{figure*}
  \centering
  \includegraphics[width=\linewidth]{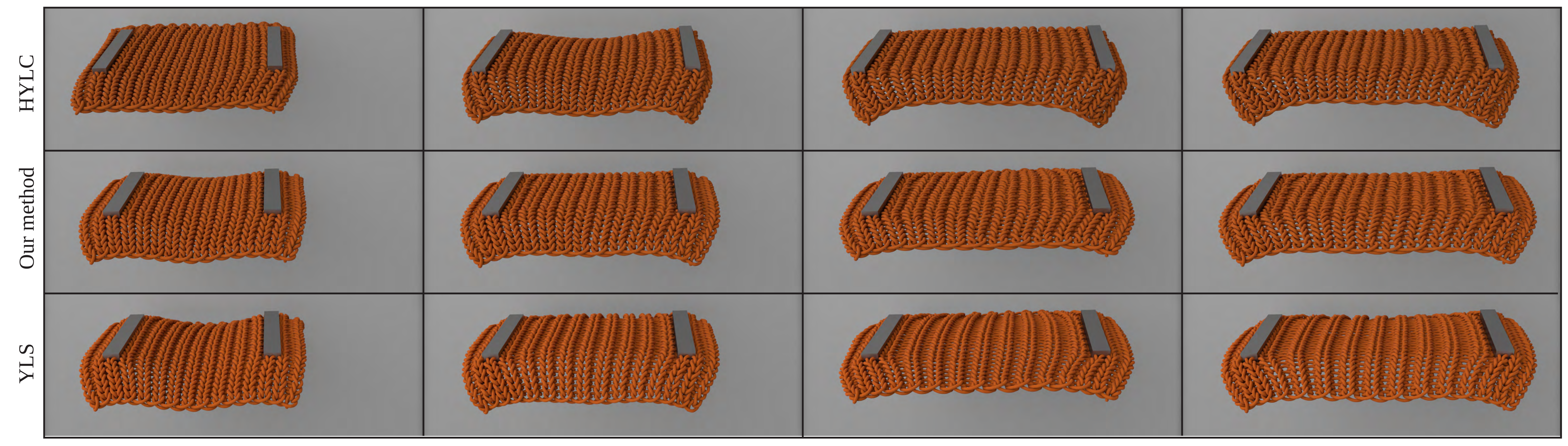}
  \caption{\textbf{Comparison with HYLC (1$\times$1 rib).}~~We compare our method with HYLC and homogenize a square knitted fabric of the periodic 1$\times$1 rib pattern. The full-scale YLS results are shown at the bottom for reference. In general, both our method and HYLC yield plausible results, and both are visually similar to YLS results. \highlight{Volumetric homogenization} uses a mesh of $38$K DOFs, while HYLC only has $9$K DOFs. Nevertheless, our method takes 85ms per frame and is $\sim 180\times$ faster than HYLC.
  }   
  \label{fig:hylc_rib}
  \Description{}
\end{figure*}

Without loss of generality, let us assume the mesh is decomposed into two domains. If we extract rows and columns corresponding to DOFs of the $d$-th domain for $d = 1, 2$, a domain-level global system can be built as
\begin{equation}
    \left[
    \begin{array}{cc}
    \bm{K}_{(d)}^{(ii)} & \bm{K}_{(d)}^{(ib)}\\
    \bm{K}_{(d)}^{(bi)} & \bm{K}_{(d)}^{(bb)}
    \end{array}
    \right]
    \left[
    \begin{array}{c}
    \bm{x}_{(d)}^{(i)}\\
    \bm{x}_{(d)}^{(b)}     
    \end{array}
    \right]
    =
    \left[
    \begin{array}{c}
    \bm{b}_{(d)}^{(i)}\\
    \bm{b}_{(d)}^{(b)}     
    \end{array}
    \right].
\end{equation}
Here, subscripts $(\cdot)^{(i)}$ and $(\cdot)^{(b)}$ denote the DOF type i.e., either internal or boundary; the subscript $(\cdot)_{(d)}$ indicates the domain index. Switching the order of subscripts applies the matrix transpose i.e., $\bm{K}_{(d)}^{(ib)} = \bm{K}_{(d)}^{(bi)^\top}$. To analyze the internal response of the domain, we prescribe a unit displacement at each boundary DOF, and pre-compute its internal response for some unknown boundary stimuli

\begin{equation}\label{eq:cms_boundary}
    \left[
    \begin{array}{cc}
    \bm{K}_{(d)}^{(ii)} & \bm{K}_{(d)}^{(ib)}\\
    \bm{K}_{(d)}^{(bi)} & \bm{K}_{(d)}^{(bb)}
    \end{array}
    \right]
    \left[
    \begin{array}{c}
    \bm{\Psi}_{(d)}^{(i)}\\
    \bm{I}     
    \end{array}
    \right]
    =
    \left[
    \begin{array}{c}
    \bm{0}\\
    \bm{F}_{(d)}^{(b)}     
    \end{array}
    \right],
\end{equation}
where $\bm{I}$ is an identity matrix corresponding to the prescribed boundary displacements, and $\bm{F}_{(d)}^{(b)}$ is the external stimuli (they are not the ``forces'' but a type of system load in a more general sense). We do not really care about the value of $\bm{F}_{(d)}^{(b)}$ but are more interested in the system response at internal DOFs, i.e., $\bm{\Psi}_{(d)}^{(i)}$. It can be computed via expanding the first row of \autoref{eq:cms_boundary}:
\begin{equation}
    \bm{K}_{(d)}^{(ii)}  \bm{\Psi}_{(d)}^{(i)} + \bm{K}_{(d)}^{(ib)} = \bm{0} \Rightarrow \bm{\Psi}_{(d)}^{(i)} = - \left(\bm{K}_{(d)}^{(ii)}\right)^{-1} \bm{K}_{(d)}^{(ib)}.
\end{equation}
Due to the linearity of this problem, $\bm{\Psi}_{(d)}^{(i)}$ encodes all the possible internal responses induced by boundary stimuli. When the domain does not undertake any non-boundary loads, $\bm{\Psi}_{(d)}^{(i)}$ relates boundary DOFs and internal DOFs as $\highlight{\bm{x}^{(i)}_{(d)} = \bm{\Psi}_{(d)}^{(i)}\bm{x}^{(b)}_{(d)}}$.
They are a.k.a. \emph{boundary modes} in CMS, and serve as subspace basis vectors for our global solve. For non-boundary responses, we compute a compact set of eigenvectors (e.g., 20) of $\bm{K}_{(d)}^{(ii)}$ corresponding to the smallest eigenvalues
\begin{equation}
\left(\bm{\Phi}^{(i)}_{(d)}\right)^\top \bm{K}_{(d)}^{(ii)} \bm{\Phi}^{(i)}_{(d)} = \bm{\Lambda}_{(d)} \;,
\end{equation}
where $\bm{\Lambda}_{(d)}$ is the diagonal matrix of eigenvalues. The composition of $\bm{\Psi}_{(d)}^{(i)}$ and $\bm{\Phi}_{(d)}^{(i)}$ constitutes a linear subspace for the internal DOFs of the domain:
\begin{equation}
    \bm{x}^{(i)}_{(d)} = \left[\bm{\Psi}_{(d)}^{(i)}, \bm{\Phi}_{(d)}^{(i)}\right]
    \left[
    \begin{array}{c}
         \bm{p}_{(d)}^{(i)}  \\
         \bm{x}_ {(d)}^{(b)}
    \end{array}
    \right].
\end{equation}

Here, $\bm{p}_{(d)}^{(i)}$ is the generalized coordinate of the $d$-th domain for its non-boundary-driven deformation. The internal boundary-driven deformation is fully prescribed by its boundary deformation of $\bm{x}_{(d)}^{(b)}$. This allows us to construct a global subspace basis matrix, which has a block-wise structure, to relate the reduced coordinate and the fullspace coordinate as
\begin{equation}\label{eq:global_subspace}
    \left[
    \begin{array}{c}
         \bm{x}_{(1)}^{(i)}  \\
         \bm{x}_{(1)}^{(b)}  \\
        \bm{x}_{(2)}^{(i)} \\
        \bm{x}_{(2)}^{(b)}
    \end{array}
    \right]
    =
    \left[
    \begin{array}{ccc}
    \bm{\Phi}_{(1)}^{(i)} & \bm{0} & \bm{\Psi}_{(1)}^{(i)} \\
    \bm{0} & \bm{0} & \bm{I}\\
    \bm{0} & \bm{\Phi}_{(2)}^{(i)} & \bm{\Psi}_{(2)}^{(i)} \\
    \bm{I} & \bm{0} & \bm{0}\\
    \end{array}
    \right]
    \left[
    \begin{array}{c}
    \bm{p}^{(i)}_{(1)} \\ 
    \bm{p}^{(i)}_{(2)} \\ 
    \bm{x}^{(b)} \\ 
    \end{array}
    \right].
\end{equation}
It is known that a domain-decomposed global system must include extra boundary constraints to make sure domains are seamlessly connected such as
\begin{align}\label{eq:dd}
   \left[
    \begin{array}{cccc}
         \bm{K}_{(1)}^{(ii)} &  \bm{K}_{(1)}^{(ib)} & \bm{0} & \bm{0}\\
         \bm{K}_{(1)}^{(bi)} &  \bm{K}_{(1)}^{(bb)} & \bm{0} & \bm{0}\\
         \bm{0} & \bm{0} & \bm{K}_{(2)}^{(ii)} & \bm{K}_{(1)}^{(ib)}\\
         \bm{0} & \bm{0} & \bm{K}_{(2)}^{(bi)} & \bm{K}_{(1)}^{(bb)}\\
    \end{array}
    \right]
    &
    \left[
    \begin{array}{c}
         \bm{x}_{(1)}^{(i)}\\
         \bm{x}_{(1)}^{(b)}\\
         \bm{x}_{(2)}^{(i)}\\
         \bm{x}_{(2)}^{(b)}
    \end{array}
    \right]
    =
    \left[
    \begin{array}{c}
         \bm{b}_{(1)}^{(i)}\\
         \bm{b}_{(1)}^{(b)}\\
         \bm{b}_{(2)}^{(i)}\\
         \bm{b}_{(2)}^{(b)}
    \end{array}
    \right]\\
 & \text{s.t.}\quad \bm{x}_{(1)}^{(i)} = \bm{x}_{(2)}^{(i)}.
\end{align}
CMS often uses the Lagrange multiplier method to enforce the boundary constraint~\cite{yang2013boundary}. Our novel subspace structure of \autoref{eq:global_subspace} eliminates duplicated boundary DOFs at the domains' interface as $\bm{x}_{(1)}^{(b)}$ and $\bm{x}_{(2)}^{(b)}$ are now uniformly written as $\bm{x}^{(b)}$. The global step of PD can be more efficiently solved by projecting \autoref{eq:dd} into the subspace of \autoref{eq:global_subspace}. The reduced global matrix remains block-sparse, and the pre-computations of \autoref{eq:cms_boundary} can be carried out in parallel at domains. 

\begin{figure*}
  \centering
  \includegraphics[width=\linewidth]{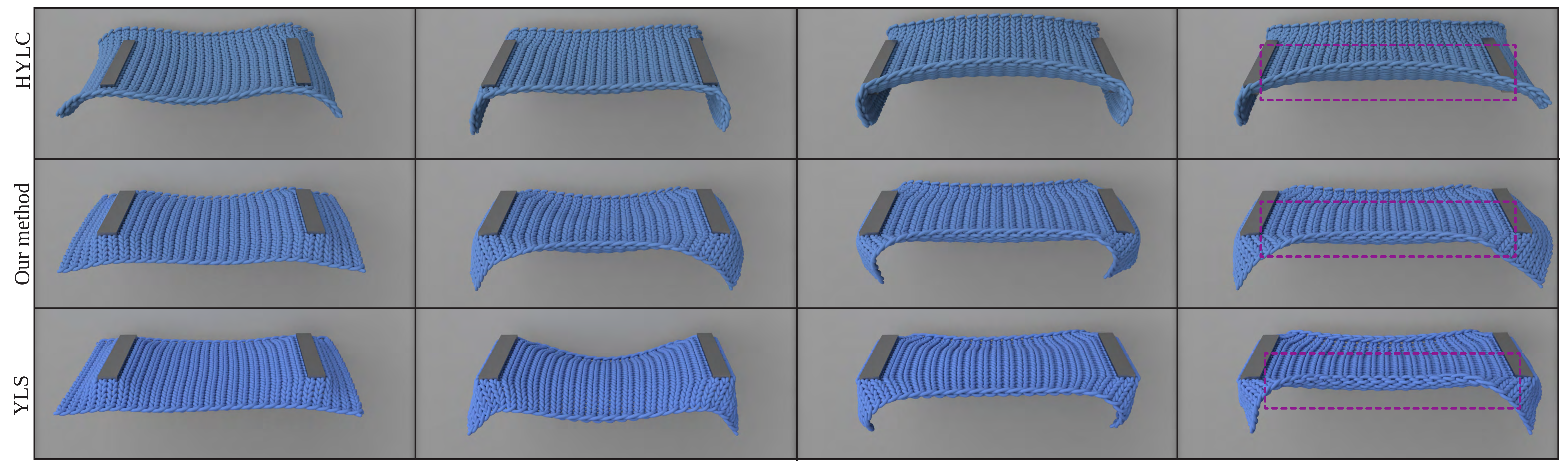}
  \caption{\textbf{Comparison with HYLC (stockinette).}~~Given fabric with a periodic stockinette pattern, our method effectively captures spatially varying deformation leveraging material heterogeneity, particularly under significant stretch as highlighted. In the example, \highlight{volumetric homogenization} uses a mesh of $42$K DOFs, while HYLC has $12$K DOFs. Our method is $\sim 350 \times$ faster than HYLC.
  }   
  \label{fig:hylc_stock}
  \Description{}
\end{figure*}
\begin{figure}
  \centering
  \includegraphics[width=\linewidth]{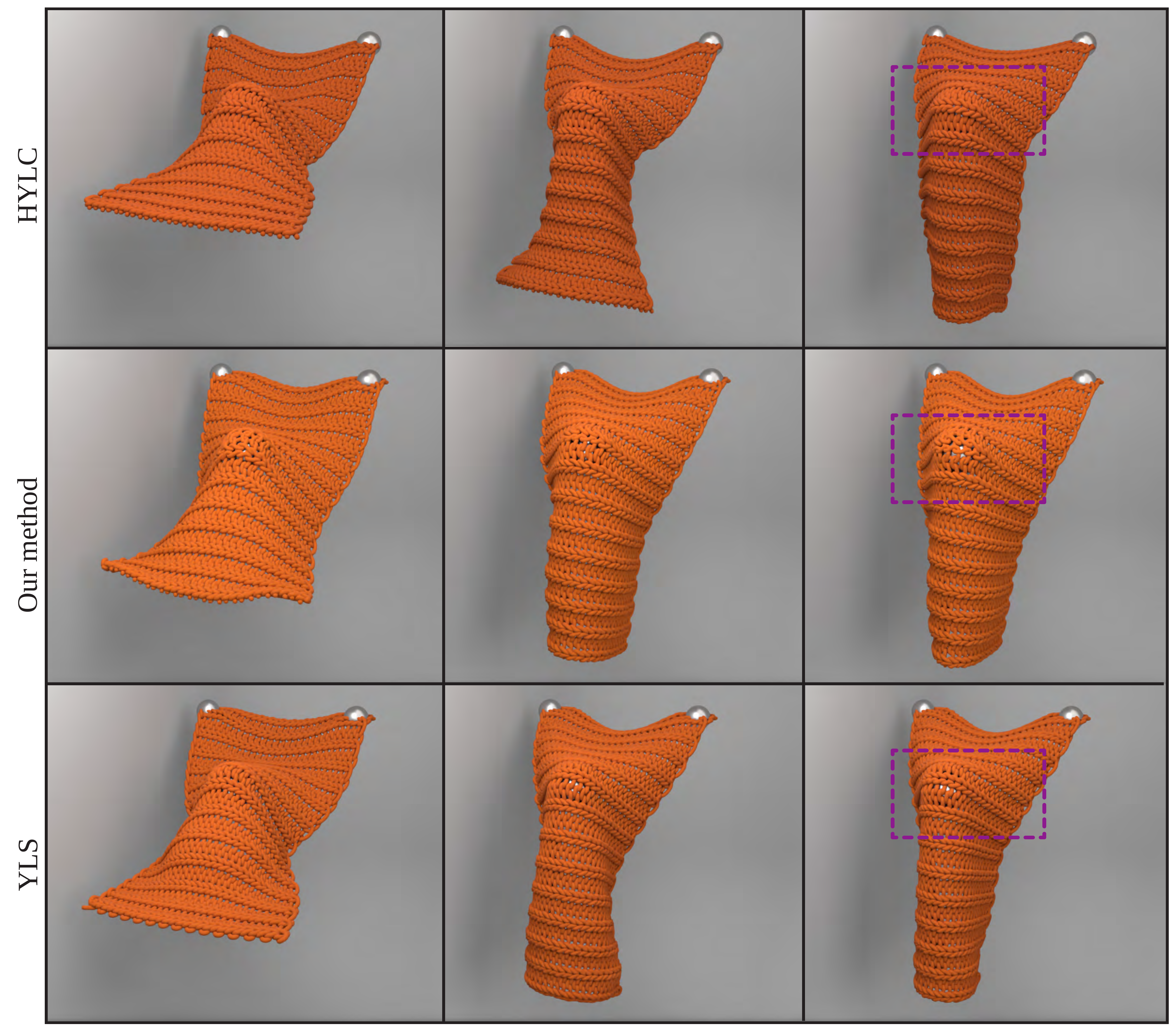}
    \caption{\textbf{Comparison with HYLC (draping).}~~We drape the \highlight{pre-stretched} ribbing patch on a sphere. Our method better replicates local deformation in the middle than HYLC. YLS simulation results are not used for elasticity fitting, showing our method's generalizability.}  
  \label{fig:hylc_drape}
  \Description{}
\end{figure}

Domain decomposition and CMS-like subspace construction techniques make our pipeline less sensitive to global matrix change e.g., due to material update or the presence of new constraints of collisions. After the reduced global solve is obtained, we use \autoref{eq:global_subspace} to convert the generalized coordinate to the fullspace result. This result is then sent to a GPU-based full-space solver to make sure the global solution is accurate. In our implementation, we use the aggregated Jacobi method (A-Jacobi) proposed in~\cite{lan2022penetration}. A-Jacobi combines two or three Jacobi iterations into one iteration to fully exploit the parallel capacity of modern GPUs. As a result, mesh-level simulation is efficient is nearly interactive in many examples reported in this paper. 

Note that a fast and numerically stable mesh simulation is critical to the entire pipeline--it is not only helpful for simulating the \highlight{volumetric homogenized} knitwear materials but also an indispensable module for elasticity fitting. Recall that our adjoint Gauss-Newton is only feasible when \autoref{eq:g} is satisfied. This requirement suggests we need to find the equilibrium configuration of the mesh given the current material vector $\bm{\gamma}$. Whenever $\bm{\gamma}$ are updated during the elasticity fitting process, we have to re-assemble and factorize $\bm{x}^M_i$, which is the most expensive computation along the pipeline. Without a fast solver, elasticity fitting at such a high-dimension material space is infeasible.

\begin{figure*}
  \centering
  \includegraphics[width=\linewidth]{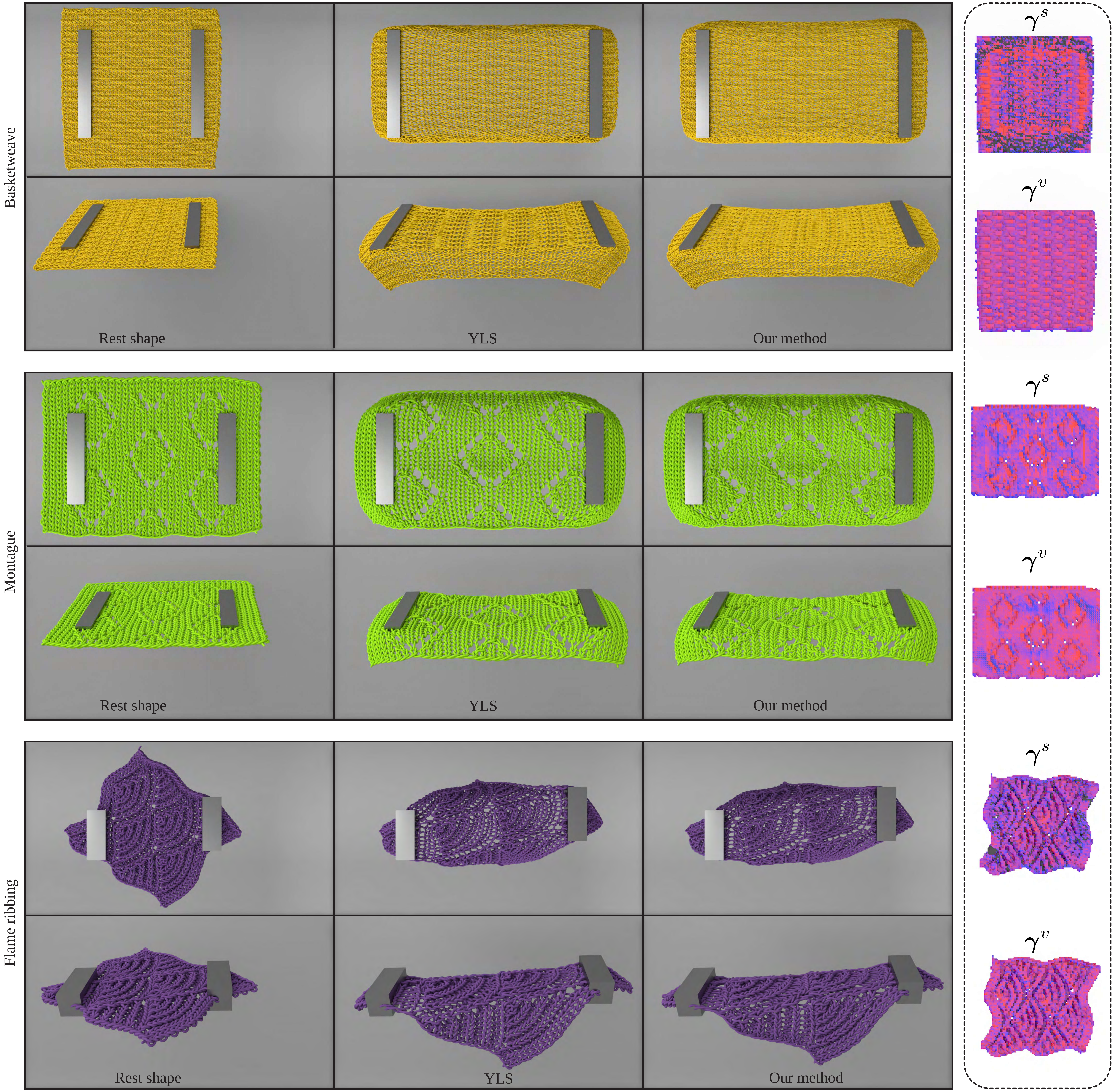}
  \caption{\textbf{More patterns.}~~We compare our results with YLS on basketweave, montague, and flame ribbing patterns. The leftmost column displays the rest shape of the knit patch, while the middle and right columns show the simulation results when the fabric is stretched using our method and YLS. We provide renderings from top and side views on the top and bottom rows, respectively. Our method accurately captures desired yarn-level deformations that vary across the patch and exhibit anisotropic behaviors. The material distributions of $\bm{\gamma}^s$ and $\bm{\gamma}^v$ are visualized on the right.}   
  \label{fig:stretch}
  \Description{}
\end{figure*}

\begin{table*}
\caption{\textbf{Statistics.}~~We report detailed time statistics for experiments mentioned in the paper. $n^Y$ is the total number DOFs on the yarn model. $n^M$ and $n^E$ are the numbers of mesh DOFs and mesh elements (and the deformable objects e.g., in \autoref{fig:puffer}). ${\# \; d}$ is the number of domains used. ${\#}$ {GN} reports the average number of adjoint Gauss-Newton iterations needed for elasticity fitting for each YLS pose. $n^F$ is the total number of sample poses used in the example. {Fit.} gives the total time used for elasticity fitting. ${\Delta t}$ is the time step size of the forward simulation, and {Sim.} gives the total time to simulate the homogenized knitwear for one time step.}\label{tab:time}
{
\begin{center}
\begin{tabular}{lc||c|c|c|c|c|c|c|c|c}
\whline{1.15pt}
 & & $n^Y$  & $n^M$  & $n^E$ & ${\# \; d}$ & ${\#}$ {GN} & ${n^F}$ & {Fit.} & ${\Delta t}$ & {Sim.} \\
\whline{0.5pt}

 \small{Comp. HYLC (1$\times$1 rib)} & \footnotesize{(\autoref{fig:hylc_rib})}  
& $66$K & $38$K & $79$K & $1$ & 15 & $2$ & $0.5$ hours & $1/150$ sec &  $85$ ms \\

 \small{Comp. HYLC (stockinette)} & \footnotesize{(\autoref{fig:hylc_stock})}  
& $66$K & $42$K & $83$K & $1$ & 42 & $3$ & $3.1$ hours & $1/150$ sec &  $92$ ms \\

 \small{Comp. HYLC (draping)}  & \footnotesize{(\autoref{fig:hylc_drape})}  
& $66$K & $42$K & $83$K & $1$ & 26 & $2$ & $1.2$ hours & $1/150$ sec &  $103$ ms \\

 \small{Basketweave } & \footnotesize{(\autoref{fig:stretch})}  
& $885$K & $107$K & $204$K & $6$ & 17 & $4$ & $2.2$ hours & $1/150$ sec &  $137$ ms \\

 \small{Montague } & \footnotesize{(\autoref{fig:stretch})}  
& $85$K & $48$K & $90$K & $9$ & 10 & $4$ & $0.7$ hours & $1/150$ sec &  $96$ ms \\

 \small{Flame } & \footnotesize{(\autoref{fig:stretch})}  
& $88$K & $60$K & $92$K & $7$ & 11 & $4$ & $0.8$ hours & $1/150$ sec &  $117$ ms \\

 \small{Mixed } & \footnotesize{(\autoref{fig:mixed})}  
& $118$K & $77$K & $151$K & $2$ & 16 & $6$ & $0.6$ hours & $1/150$ sec &  $121$ ms \\

 \small{Make a knot } & \footnotesize{(\autoref{fig:knot})} 
& $277$K & $78$K & $91$K & $4$ & 37 & $5$ & $4.2$ hours & $1/200$ sec &  $335$ ms \\

 \small{Twisting } & \footnotesize{(\autoref{fig:twist})}  
& $885$K & $107$K & $204$K & $6$ & 16 & $4$ & $3.7$ hours & $1/200$ sec &  $672$ ms \\

 \small{Puffer ball on the knit } & \footnotesize{(\autoref{fig:puffer})}  
& $31$K & $120$K($207$K) & $140$K($414$K) & 1 & $13$ & 2 & $0.6$ hours & $1/250$ sec &  $1132$ ms \\

 \small{Yangge dance } & \footnotesize{(\autoref{fig:teaser})}  
& $3.0$M & $342$K & $389$K & $14$ & 47 & $10$ & $41.2$ hours & $1/150$ sec &  $604$ ms \\

 \small{Short-sleeve Yangge } & \footnotesize{(\autoref{fig:garment})}  
& $2.7$M & $329$K & $361$K & $10$ & 36 & $10$ & $35.3$ hours & $1/150$ sec &  $588$ ms \\

 \small{Long-sleeve Yangge } & \footnotesize{(\autoref{fig:garment})}  
& $2.6$M & $493$K & $420$K & $8$ & 52 & $10$ & $65.0$ hours & $1/150$ sec &  $872$ ms \\

\whline{1.15pt}
\end{tabular}
\end{center}
}
\end{table*}

\section{Experimental Results}\label{sec:exp}
We implemented our \highlight{volumetric homogenization} framework on a desktop computer with an 
Intel i7-12700 CPU and an NVIDIA RTX 3090 GPU.
We used \texttt{Spectra} library for computing the eigendecomposition of the mesh Laplacian, and $\bm{K}^{ii}_{d}$ for each domain, and \texttt{Eigen}~\cite{guennebaud2010eigen} as the primary interface of linear algebra computations. The domain-decomposed PD procedure consists of two steps. The first step solves the global matrix in the subspace, and we employed the \texttt{PARDISO} solver~\cite{schenk2001pardiso} shipped with Intel \texttt{MKL}~\cite{wang2014intel}. Fullspace A-Jacobi iterations~\cite{lan2022penetration} are followed to make sure the global step system is well solved. This step was implemented on the GPU with \texttt{CUDA}. We used Chebyshev weight to improve the convergence of A-Jacobi method~\cite{Wang2015Chebyshev}. The local step is parallelized on the GPU at each element with \texttt{CUDA}. Adjoint gradient descent and adjoint Gauss-Newton are sensitive to the residual of \autoref{eq:g} i.e., \autoref{eq:g} must be strictly satisfied. Therefore, during the elasticity fitting, we applied a few (three to five) Newton iterations after the PD solve to keep the residual smaller than $1E-5$. We generate full-scale YLS results by simulating each yarn as the Cosserat rod model~\cite{spillmann2007corde}. The detailed statistics of experiments are reported in \autoref{tab:time}. 

\subsection{Comparison with HYLC}
Homogenized yarn-level cloth (HYLC)~\cite{sperl2020homogenized} provides an excellent paradigm by homogenizing the ``yarn material'' to the mid-surface of the fabric, represented as a triangle mesh. Our \highlight{volumetric homogenization} method expands the dimensionality to a 3D volume with more DOFs and optimizes spatially varying material properties instead of assuming a uniform material as in HYLC. Our first experiment involves three side-by-side comparisons to thoroughly understand the differences between our method and HYLC~\footnote{We used the implementation of HYLC at~\url{https://git.ista.ac.at/gsperl/HYLC}}. \highlight{In HYLC, the Discrete Elastic Rod energy is replaced by Cosserat Rod energy for our comparison. The visualization of both our method and HYLC is achieved by embedding the yarn geometry into the mesh.}

As shown in \autoref{fig:hylc_rib}, both HYLC and our method yield visually similar and plausible results for a knitted patch with a 1$\times$1 rib pattern compared to YLS as the ground truth. The 1$\times$1 rib pattern is highly stretchable due to the alternating knit and purl columns, allowing the fabric to expand and contract significantly across its width. However, HYLC, utilizing a single material across the entire mesh, causes the patch to narrow in the perpendicular directions when stretched, exhibiting a rubber-like behavior (see the rightmost column in \autoref{fig:hylc_rib}).
The inconsistency between our method and HYLC becomes more apparent for the stockinette patch, which is less stretchy than the rib but tends to curl at the edges. While both sheet-based (HYLC) and volume-based (ours) materials display curly edges under stretch, HYLC is less expressive in capturing subtle deformation variations across the fabric due to the lack of material heterogeneity. In contrast, our method better replicates this phenomenon.
We further test by draping a \highlight{pre-stretched} 1$\times$1 rib patch over a sphere, demonstrating a similar difference to that shown in \autoref{fig:hylc_stock}. Both HYLC and our method produce reasonable global deformation. However, our method, with a spatially varying material, better captures the inhomogeneous deformation at the contacting region. In this set of experiments, only two or three $\bm{x}^Y_i$ are used for elasticity fitting, and the draping simulation results are not observed during the elasticity fitting.

In terms of performance, the tetrahedron mesh used in our method consists of approximately $80$K elements. This is a bigger number compared with HYLC, which simulates a triangle mesh of around $10$K triangles. However, our constraint-based material model (\autoref{eq:energy}) allows for a more stable simulation with a larger time step. The domain-decomposed PD solver completes these experiments at $\Delta t = 1/150$ sec, while HYLC must reduce to $\Delta t = 1/500$ sec (or even smaller) to ensure stability and prevent divergence due to material nonlinearity. Additionally, HYLC adaptively subdivides the mesh using \texttt{ArcSim}~\cite{narain2013folding} to improve stability. Overall, our method is two orders faster than HYLC. The total training time is 0.5 hours, 3.1 hours, and 1.2 hours for \autoref{fig:hylc_rib}, \autoref{fig:hylc_stock}, and \autoref{fig:hylc_drape}, respectively.

\begin{figure*}
  \centering
  \includegraphics[width=\linewidth]{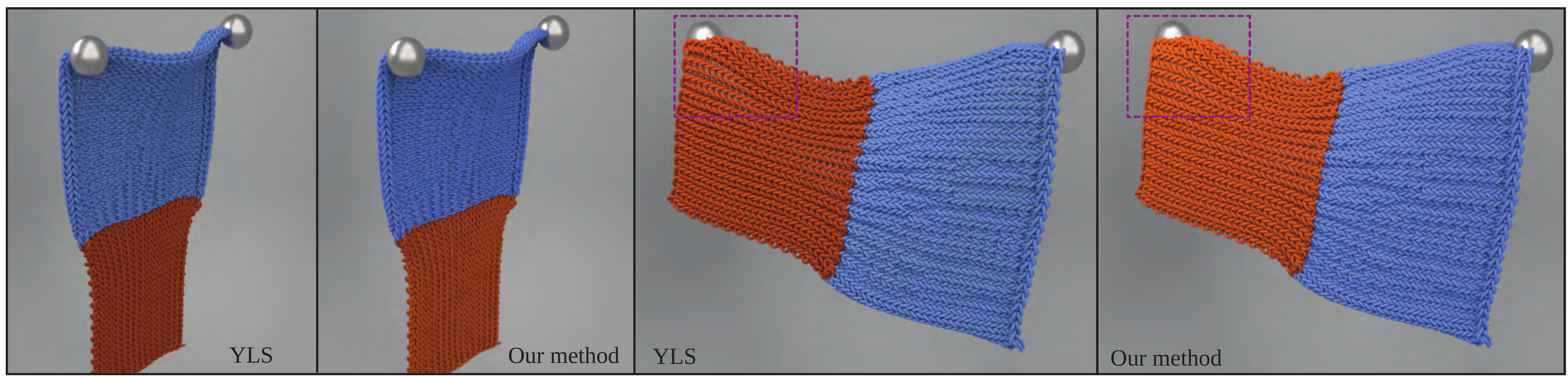}
  \caption{\textbf{Mixed patterns.}~~Our domain decomposed forward simulation is naturally compatible with garments made of combined patterns. In this example, we simulate a hanging fabric consisting of two patterns: knit and rib. One can still observe nuanced differences between our method and YLS, e.g., see highlighted areas. Nevertheless, homogenized simulation yields good animation results.}   
  \label{fig:mixed}
  \Description{}
\end{figure*}
\subsection{More patterns}
We further compare our results with yarn-level simulation (YLS) using three additional knit patterns: basketweave, montague, and flame ribbing. Particularly, flame ribbing patterns exhibit anisotropic spatial variation with non-periodic yarn structures, making it challenging to fit a homogeneous material model using HYLC. In contrast, our \highlight{volumetric homogenization} does not rely on any assumptions about the underlying arrangement of yarn structures. As shown in~\autoref{fig:stretch}, our method produces high-fidelity knit stretching results that are nearly identical to YLS. As visualized on the right, the fitted material distributions align well with the underlying yarn structure and capture location-dependent deformations as expected.

\autoref{fig:mixed} presents another experiment featuring a fabric composed of two distinct patterns: stockinette and 1$\times$1 rib. With the same number of stitches per row, the rib pattern shrinks due to the alternating knit and purl stitches, making it highly elastic and capable of significant expansion and contraction, while the stockinette is smooth and flat. This combined pattern poses a challenge to classic homogenization theory. However, our method remains effective in this case. Our domain-decomposed forward simulator naturally accommodates such pattern combinations. While YLS provides richer local details in this example, our method still produces reasonably good results. In this case, we have $77$K DOFs on the mesh, and the simulation takes $121$ ms for one time step, which is $\sim 80 \times$ faster than running the simulation at the yarn level.

\begin{figure}
  \centering
  \includegraphics[width=0.9\linewidth]{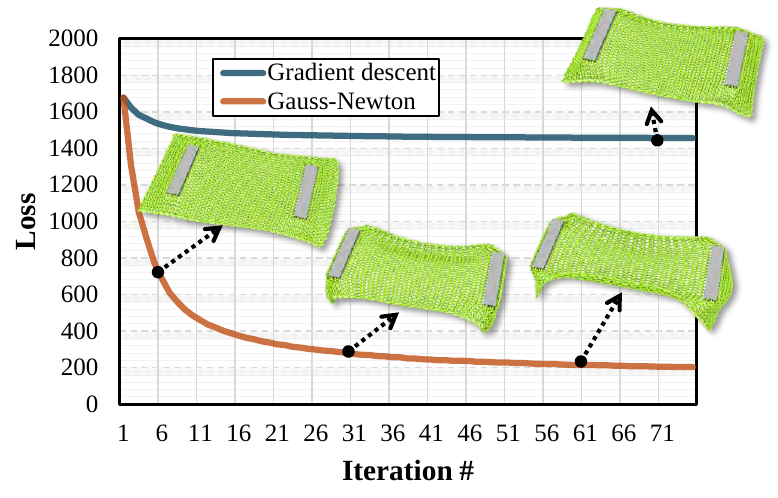}
  \caption{\textbf{Convergence plots.}~~We plot the converge curves using adjoint Gauss-Newton and gradient descent. The first-order method does not only converge properly, while second-order information in $\bm{G}_i$ (\autoref{eq:agn}) effectively helps find a better descent direction and lower the loss function.}   
  \label{fig:convergence}
  \Description{}
\end{figure}

\subsection{Convergence}
The key to a successful \highlight{volumetric homogenization} is to solve the inverse problem using high-order optimization techniques i.e., the adjoint Gauss-Newton method discussed in \autoref{sec:elasticity_fitting}. We note that most existing gradient-based methods, either using adjoint method or using AD (automatic differentiation) fail to converge in our case. We plot the convergence curves using adjoint Gauss-Newton and gradient descent for a representative elasticity fitting instance and show the result in \autoref{fig:convergence}. We also visualize the deformed mesh that satisfies \autoref{eq:g} at three different iterations. A consistent observation is that first-order methods are never going to work. They frequently get trapped at local minima. A more severe issue is the step size of the gradient-based method is not stable. Performing line search is expensive in elasticity fitting -- any proposed material update $\Delta \bm{\gamma}$ can only be checked when \autoref{eq:g} is satisfied, and therefore a forward simulation procedure is invoked. This makes the first-order method prohibitive in practice. For instance, it could take over one week if we choose to use the first-order adjoint method or differentiable simulation to optimize the rib pattern (\autoref{fig:hylc_rib}). On the other hand, Gauss-Newton finishes the training in about three hours.

\begin{figure}
  \centering
  \includegraphics[width=\linewidth]{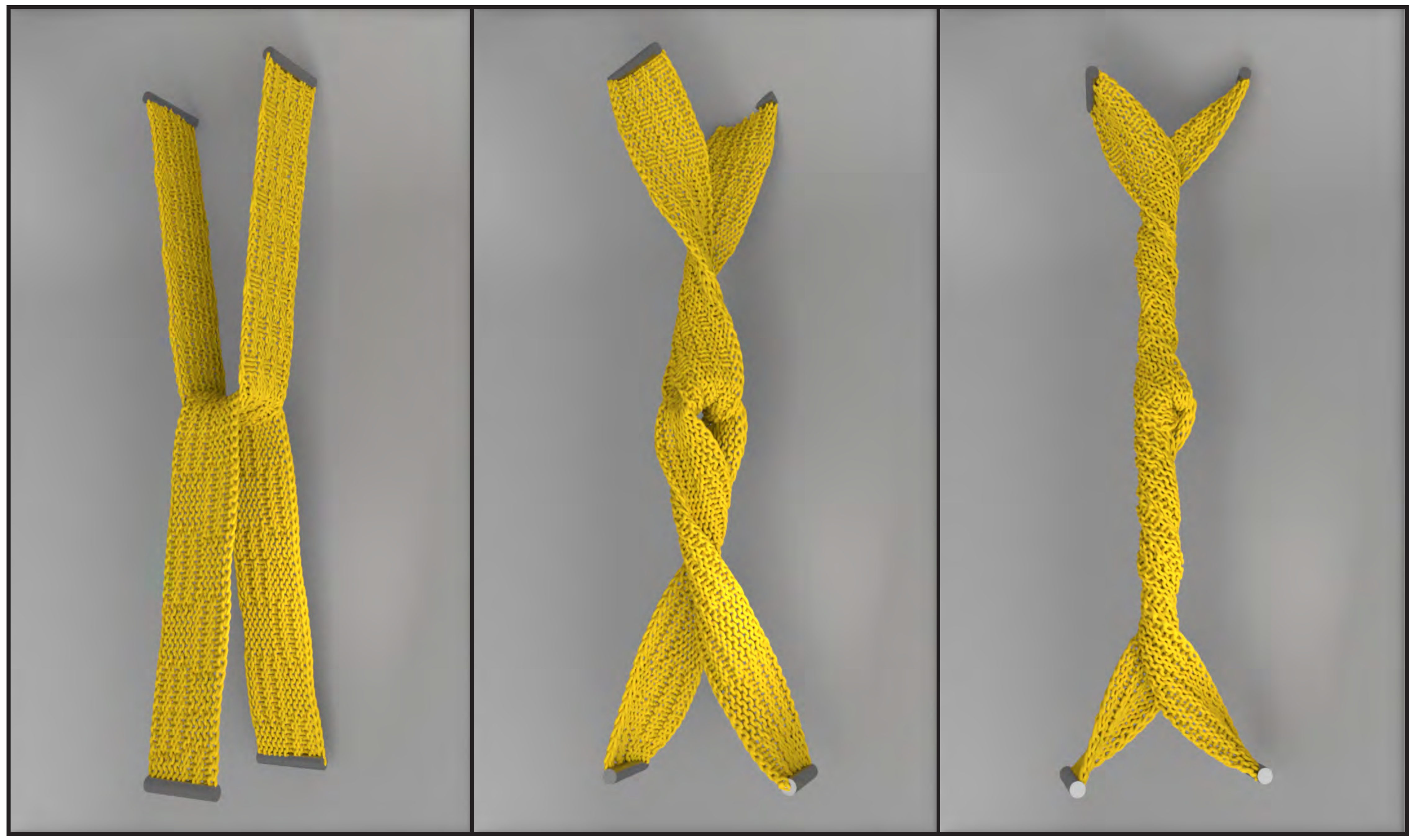}
  \caption{\textbf{Make a knot.}~~We rotate two knitted scripts (garter block pattern for the top and basketweave pattern for the bottom) for $1,920$ degrees to make a knot. There are $78K$ DOFs on the mesh, and homogenized simulation takes 335 ms for each time step ($\Delta t = 1/200$).}   
  \label{fig:knot}
  \Description{}
\end{figure}

\begin{figure*}
  \centering
  \includegraphics[width=\linewidth]{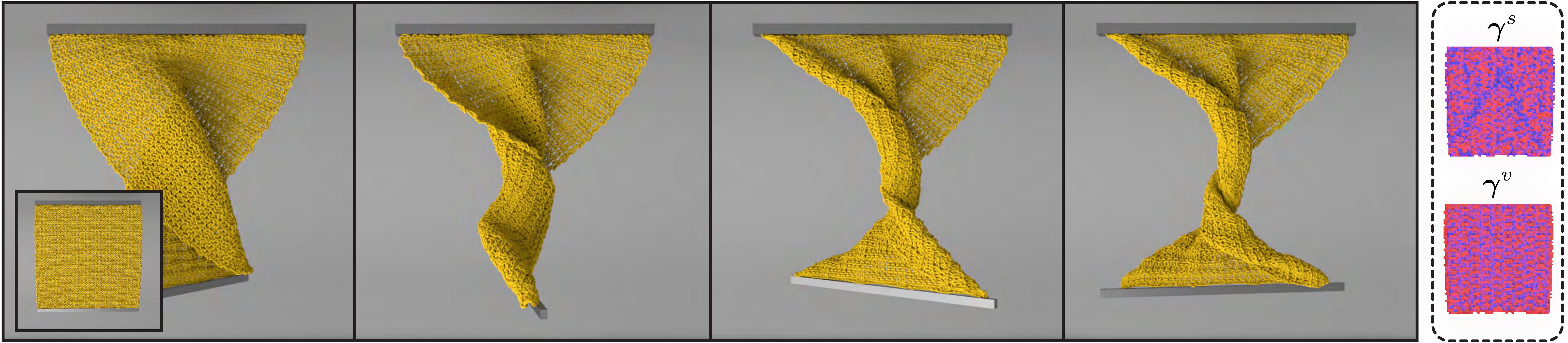}
  \caption{\textbf{Twisting.}~~We twist a square basketweave fabric for ~ $900$ degrees. In this example, the final deformed shape of our method does not perfectly align with the YLS result (please refer to the supplementary video for the animated simulation result). This is because the shape discrepancy accumulates at each frame, leading to different fabric-fabric collisions during the twisting. While not exactly same as YLC, our results are natural and realistic. There are $107$K DOFs on the mesh. The homogenized simulation takes $672$ ms for each time step ($\Delta t = 1/200$) and is $\sim 120 \times$ faster.}   
  \label{fig:twist}
  \Description{}
\end{figure*}

\begin{figure*}
  \centering
  \includegraphics[width=\linewidth]{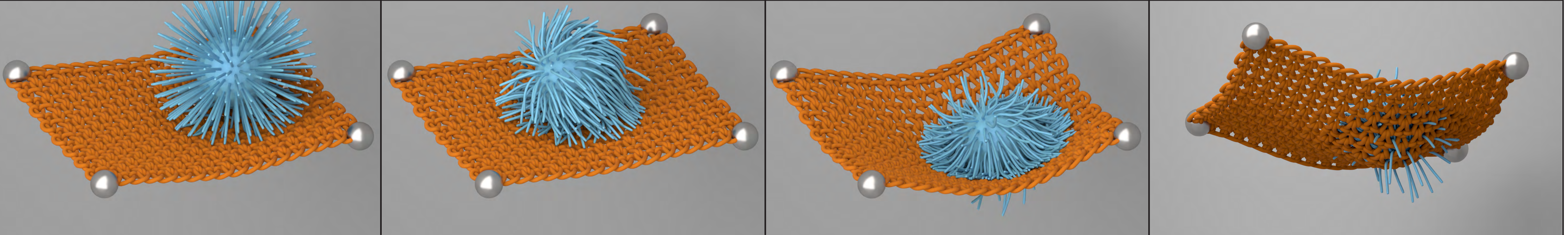}
  \caption{\textbf{Puffer ball on the knit.}~~We separate the processing of energy constraint (on the mesh) and collision constraint (at the yarn level) so that the homogenized model interacts as a yarn-level model. In this example, a puffer ball falls on a stockinette fabric. The hairs on the puffer ball pass through the gaps on the knit. They are two-way coupled under collision constraints and can be conveniently processed with our simulation.}   
  \label{fig:puffer}
  \Description{}
\end{figure*}

\begin{figure*}
  \centering
  \includegraphics[width=\linewidth]{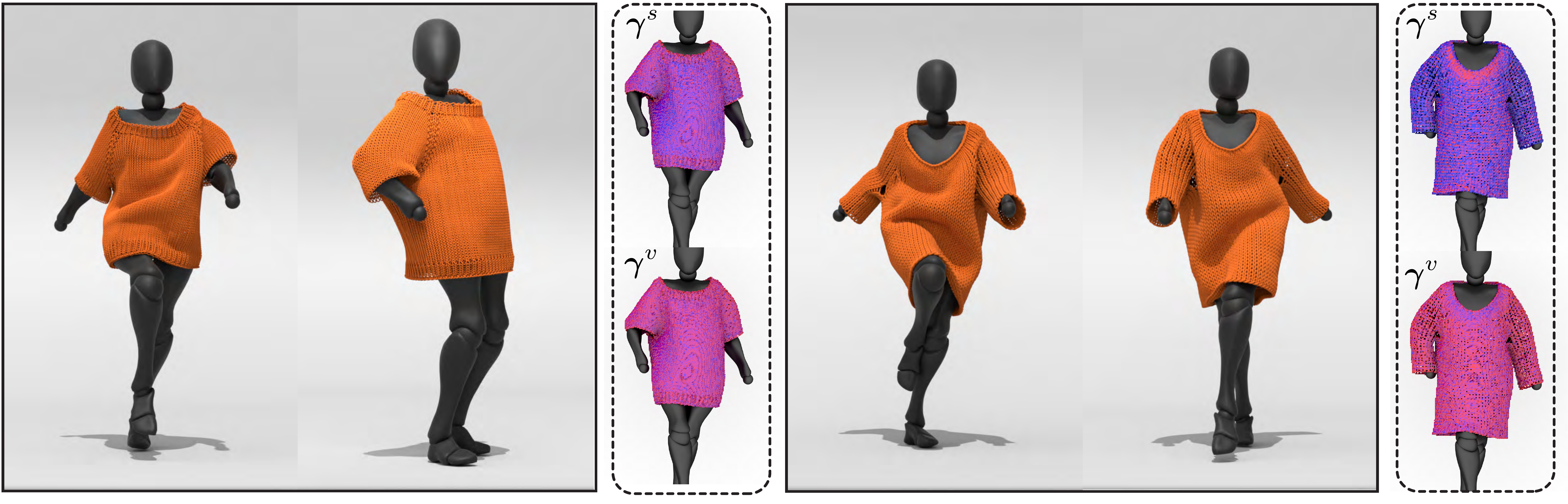}
  \caption{\textbf{Short- and long-sleeve Yangge.}~~In addition to \autoref{fig:teaser}, we show two more examples of full garment animations. On the left, the character wears a knitted short-sleeve sweater. It has a different pattern (stockinette pattern) than the one shown in \autoref{fig:teaser}, which yields different garment dynamics. On the right, the character wears a long-sleeve sweater in a stockinette pattern. The material distributions are visualized on the right.}   
  \label{fig:garment}
  \Description{}
\end{figure*}

\subsection{\highlight{Comparision at Different Resolution}}
\begin{figure*}
  \centering
  \includegraphics[width=\linewidth]{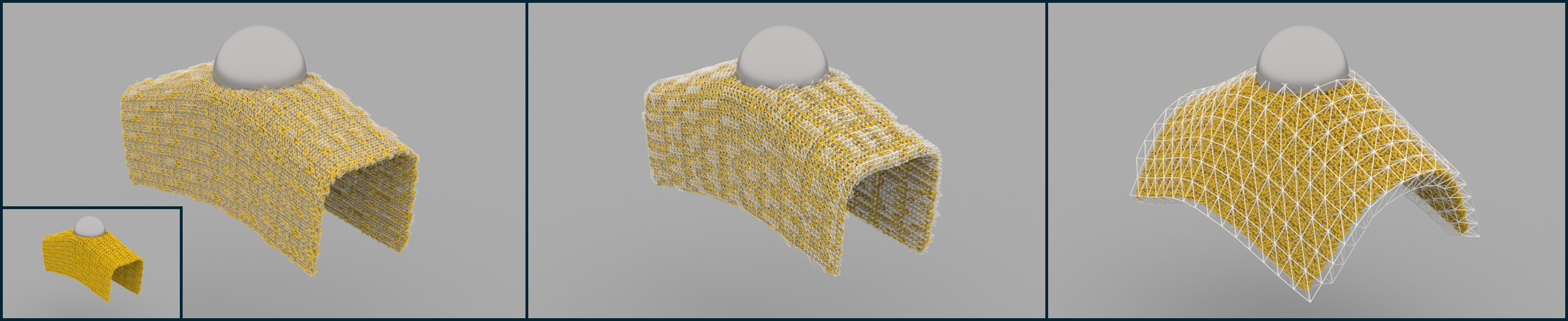}
  \caption{\textbf{Comparison at different resolution.}~~\highlight{We compare our method's fitting capabilities across varying mesh resolutions by presenting an instance of a basketweave fabric draped over a sphere. Each panel displays the fabric fitting result at a different level of resolution. The smallest thumbnail inset in the bottom left panel provides the ground truth. The middle and the leftmost panels, where each cell's size is similar to or smaller than the size of each loop structure of threads, show fitting results that are essentially consistent with the ground truth. However, the rightmost panel, with a much coaser resolution, displays anisotropic bending effects macroscopically but significantly loses sharp bending details.}}   
  \label{fig:comp-resolution}
  \Description{}
\end{figure*}

\highlight{The resolution of our volumetric material significantly influences the fitting capability of our method. We compare our method's performance on basketweave material at various resolutions as shown in \autoref{fig:comp-resolution}. The center of the basketweave patch was fixed with a sphere. While the boundary condition exhibits rotational symmetry, the basketweave demonstrates distinct bending behaviors between the course and wale directions. When the mesh resolution is close to or higher than that of the knit pattern, our approach produces results nearly identical to the ground truth. When the mesh resolution is significantly lower than that of the knit pattern, we still observe anisotropic bending behavior, but the distinction between the bending behaviors in the course and wale directions is not as sharp as in the ground truth. 
}

\subsection{More Results}
We test our method in several complex scenes. Figs.~\ref{fig:knot} and~\ref{fig:twist} report two challenging simulation scenarios involving intensive twisting and bending. In \autoref{fig:knot}, we rotate two fabric strips (basketweave) for about $1,920$ degrees. They are tightly intertwined to form a knot. In this case, \highlight{volumetric homogenization} yields a stronger volume-preserving penalty than examples shown in \autoref{fig:stretch}. We would like to mention that while the homogenized simulation uses the tetrahedron mesh for energy integration and elasticity constraint projection, we can still process collision at yarn segments if needed (e.g., see \autoref{fig:puffer}). \highlight{As most yarn-yarn contacts are inelastic, we follow the collision projection strategy as in vanilla PD\mbox{~\cite{bouaziz2023projective}} for each edge-edge collision. The collision impulse is then interpolated to the corresponding tetrahedron~\cite{lei2022abd}. For collision detection, to enhance performance, we adopt the DCD (Discrete collision detection) strategy, noting that edge-edge collision pairs that exist in the same cell or adjacent cells can be excluded in advance. Of course, if we aim for strict non-penetration, we can adopt CCD (Continuous collision detection) with the interior point method as in ~\cite{li2021cipc}. Although in our experiments we used the DCD collision detection method, which does not have guarantees and may eventually result in inter-penetration, we find that our homogenized animations remain visually plausible.} \autoref{fig:twist} showcases an animation of twisting a wider fabric. Similar to \autoref{fig:knot}, our method faithfully captures the twisting behaviors of such a complicated yarn structure using our volume-preserving energy. The YLS result can be found in the supplementary video. There are $91$K elements in \autoref{fig:knot} and $204$K elements in \autoref{fig:twist}. The simulation uses $335$ ms for one time step on average in \autoref{fig:knot} and $672$ ms in \autoref{fig:twist}.

\autoref{fig:garment} and \autoref{fig:teaser} show the animation of three different knit garments on a virtual character performing Yangge dance. Different patterns on the front panels of the garments lead to interesting and distinctive animation effects even under the same character motion. Our homogenized simulations well replicate the YLS results but are two orders faster. In these three examples, we use ten poses for the elasticity training. The resulting material distributions are also visualized in the figures.

\subsection{\highlight{Generalizability}}
\begin{figure}
  \centering
  \includegraphics[width=\linewidth]{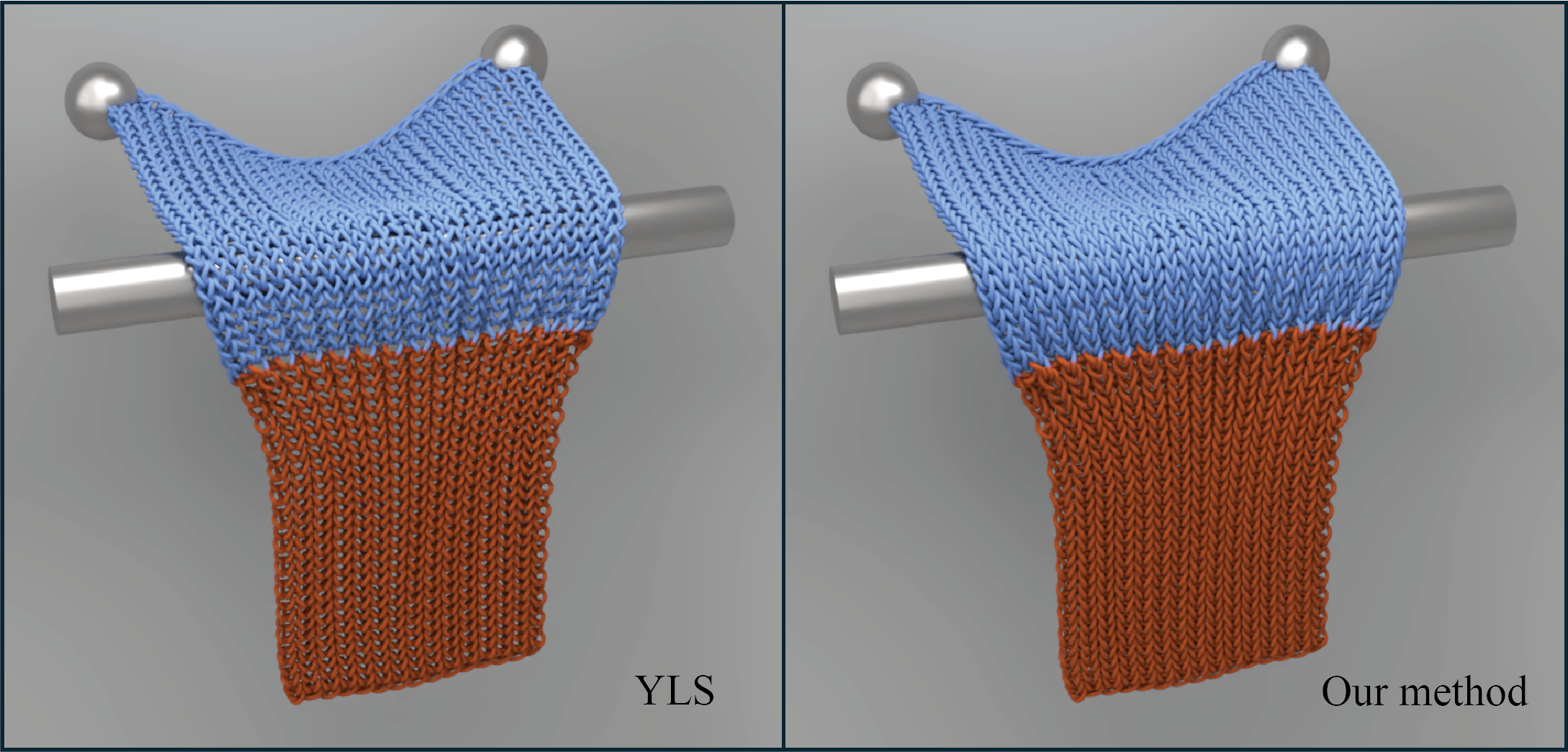}
  \caption{\textbf{Mixed patterns (draping).}~~\highlight{We drape the mixed pattern on a cylinder, reusing the material of the mixed pattern trained in \autoref{fig:mixed} Our method exhibits almost the same behavior as YLS.}}   
  \label{fig:mix2cylinder}
  \Description{}
\end{figure}

\begin{figure*}
  \centering
  \includegraphics[width=\linewidth]{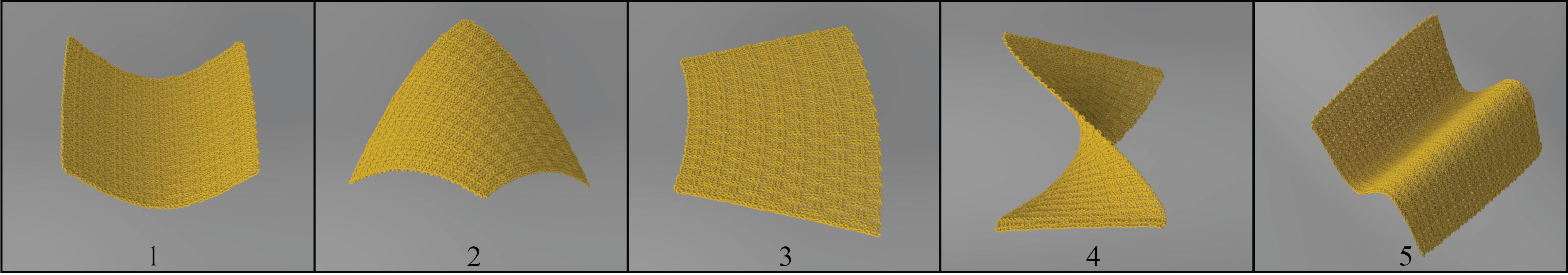} 
  \caption{\textbf{More poses.}~~\highlight{We follow the data generation protocol outlined in HYLC to sample five types of deformation, including their symmetric counterparts. The first two types are bending along the central axis and diagonal axis. The third deformation is a fan-like expansion. For the fourth type, the fabric is subject to a twist that forms a helix, we include the multi-folded deformation featuring several folds along the material.}}
  \label{fig:poses}
  \Description{}
\end{figure*}

\begin{figure*}
  \centering
  \includegraphics[width=\linewidth]{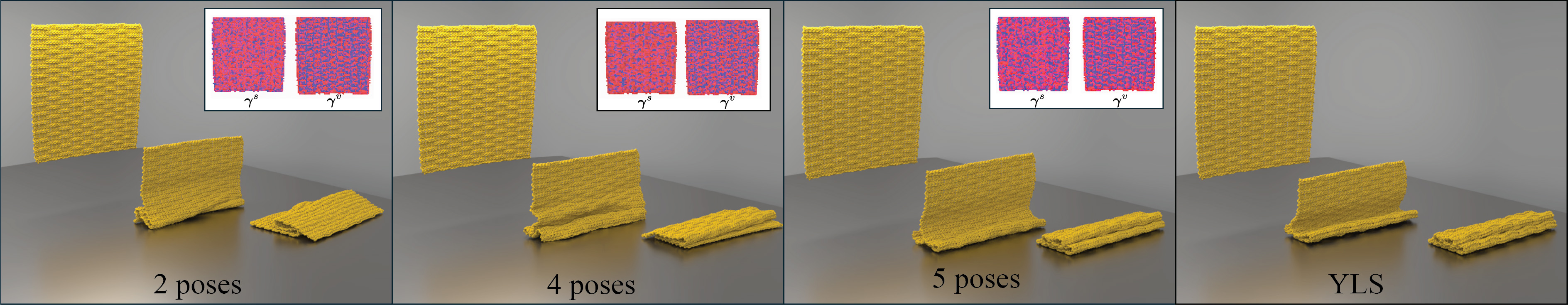}
  \caption{\textbf{Dropping to ground.}~~\highlight{We utilize the training data generated from \autoref{fig:poses} and compare the dropping behavior with different training sets. Initially, only the first two bending types are used as our training data. In the subsequent panel, the fan-like expansion and twist types are also included. All deformation types in \autoref{fig:poses} are added in the third panel. The corresponding material distributions are shown in the top-right corner of each panel. The YLS is displayed in the rightmost column.}}   
  \label{fig:to_ground}
  \Description{}
\end{figure*}

\highlight{Most of our experiment's training data is directly sampled from the animation sequence. However, in \autoref{fig:hylc_drape}, the draping test of the pre-stretched patch uses training data based on lateral stretching deformation. The final result shows a similar draping behavior to YLS, demonstrating the generalizability of our method to some extent. Here, we provide additional experiments to further explore the generalizability of our method. \autoref{fig:mix2cylinder} demonstrates the draping of mixed patterns on a cylindrical form, using the material of the mixed pattern trained as shown in \autoref{fig:mixed}, the homogenized simulation exhibits almost the same behavior. Besides sample training data from the animation sequence, we can also follow the data generation protocol as in HYLC to sample possible deformations. In \autoref{fig:poses}, we generate five types of deformation poses: Panel 1 shows a simple upward bend along the central axis, creating a smooth concave shape; Panel 2 presents a diagonal bend, where the material curves along the diagonal axis; Panel 3 illustrates a fan-like stretch configuration; Panel 4 depicts a twisting deformation that forms a helicoidal structure; and Panel 5 features a more complex, multi-folded bend with several folds along the material. Each of these poses, along with their symmetric counterparts, progressively composes the training set used to train the material distribution, as shown in \autoref{fig:to_ground}. As the training data set increases, the behavior becomes closer to that of YLS. It is important to note that the training dataset does not include any data from the YLS animation sequence shown in \autoref{fig:to_ground}.}

\section{Conclusion \& limitation}\label{sec:conclusion}
This paper explores a different perspective on improving the simulation of knitwear with complex knitting patterns. We name our method \highlight{volumetric homogenization} because it enables a volumetric and spatially heterogeneous material synthesis. The advantages of this strategy are multifold. Volume materials implicitly handle the bending and twisting and reduce the nonlinearity of the material property. It allows a faster simulation algorithm and, in turn, benefits the fitting efficiency. Assigning each volume element an independent set of material parameters effectively enhances the versatility of simulation so that our method is able to capture subtle and visually pleasing local deformations of complex knittings. With a domain-decomposed PD solver, our method is orders of magnitude faster than full-scale YLS.

While \highlight{volumetric homogenization} shows make some non-trivial improvements over existing methods, it still has drawbacks and limitations. We leverage shape fitting to extract the acceleration of YLS sequences and to isolate the inertia effects during the material learning. For fast-moving scenes however, the estimated inertia and lumped mass may still differ from the reference. Yarn-level cloth models are often highly dampened. This is because the friction and contacts among yarn threads dissipate inertia energy quickly. It is challenging to calibrate a homogenized garment with full-scale YLS using commonly seen macroscopic damping models e.g., Rayleigh damping. Being a data-driven method, our method could generate different results of the same knitwear given different YLS inputs. It is possible to learn a more sophisticated strain-stress model as in~\cite{sperl2020homogenized} at each element. Doing so requires significant computations, and the resulting material becomes strongly nonlinear (again). In other words, it remains an open question to find the right compromise between material nonlinearity, the level of visual realism, and computational efficiency.


\bibliographystyle{ACM-Reference-Format}
\bibliography{bibliography.bib}

\end{document}